\documentclass{aa}  

\usepackage{graphicx}
\usepackage{txfonts}
\usepackage{lscape}
\usepackage{hyperref}
\usepackage{placeins}
\begin{document} 

   \title{Water Ortho-to-Para ratio in the coma of comet 67P/Churyumov-Gerasimenko}

   \author{
   Y.-C. Cheng,
          \inst{1,2}
          D. Bockel\'ee-Morvan,
          \inst{1}
          M. Roos-Serote,
          \inst{1}
          J. Crovisier,
          \inst{1}
          V. Debout,
          \inst{1}
          S. Erard,
          \inst{1}
          P. Drossart,
          \inst{1}
          C. Leyrat,
          \inst{1}
           F. Capaccioni,
          \inst{3}
           G. Filacchione,
          \inst{3}
          \and
         M.-L. Dubernet
         \inst{4}
         \and 
         T. Encrenaz
         \inst{1}         
          }

   \institute{LESIA, Observatoire de Paris, Universit\'e PSL, CNRS, Sorbonne Universit\'e, Universit\'e de Paris, 
5 place Jules Janssen, 92195, Meudon Cedex, France\\
              \email{dominique.bockelee@obspm.fr}
          \and  
          Department of Physics, National Taiwan Normal University, No. 88, Section 4, Tingzhou Road, Wenshan District, Taipei City, 116325 Taiwan  
         \and
         Istituto di Astrofisica e Planetologia Spaziali, Istituto Nazionale di Astrofisica, via del Fosso del Cavaliere 100, 00133 Rome, Italy 
         \and
             LERMA, Observatoire de Paris, Universit\'e PSL, CNRS, Sorbonne Universit\'e, 5 place Jules Janssen, 92195, Meudon Cedex, France\\             
             }

   \date{Received ; accepted}

 
  \abstract
   {Abundance ratios of the nuclear-spin isomers of H$_2$O and NH$_3$ have been measured in about two dozen comets, with a mean value corresponding to a nuclear-spin temperature of $\sim$ 30 K. The real meaning of these unequilibrated nuclear-spin abundance ratios is still debated. However, an equilibrated water ortho-to-para ratio of 3 is also commonly observed.}
   {The H channel of the Visible and Infrared Thermal Imaging Spectrometer (VIRTIS-H) on board Rosetta provided high-resolution 2.5--2.9 $\mu$m spectra of H$_2$O vapour in the coma of comet 67P/Churyumov-Gerasimenko (67P), which are suitable for the determination of the ortho-to-para ratio (OPR) of water in this comet.}
   {A large dataset of VIRTIS-H spectra obtained in limb-sounding viewing geometry was analysed, covering heliocentric distances from 1.24 to 2.73 au and altitudes from a few hundred metres to $>$ 100 km. The OPR, together with the H$_2$O rotational temperature and column density, were derived for each spectra using a database of fluorescence synthetic spectra that include both fundamental and hot vibrational water bands. The weak lines of the $\nu_1$,  $\nu_1+\nu_3-\nu_1$ and $\nu_2+\nu_3-\nu_2$ bands in the 2.774--2.910 $\mu$m range were used to calculate by how much the strong $\nu_3$ band centred at 2.67 $\mu$m is attenuated due to optical depth effects, expressed by the attenuation factor $f_{\rm atten}$.}
   {Most ortho-to-para ratio determinations are strongly affected by opacity effects, as demonstrated by the observed anti-correlation between the OPR and the column density, and the correlation between the OPR and attenuation factor $f_{\rm atten}$. Based on both radiative transfer calculations and OPR values obtained in low-opacity conditions, we derive an OPR of 2.94 $\pm$ 0.06 for comet 67P. Measured water rotational temperatures show a decrease in gas kinetic temperature with increasing altitude caused by adiabatic cooling. Heliocentric variations are also observed, with warmer temperatures near perihelion.    }
   {The water ortho-to-para ratio measured in the coma of 67P is consistent with laboratory experiments showing that water vapour that has thermally desorbed from water ice has a statistical value of 3, regardless of the past formation process of water ice.}

\titlerunning{Water OPR in comet 67P}
\authorrunning{Cheng et al.}
   \maketitle
%

\section{Introduction}

Comets are providing interesting clues regarding the history of the Solar System.
Through their mineralogical, chemical, and isotopic composition, comet nuclei document 
environmental conditions and processes occurring from the protostellar collapse phase to the protoplanetary disk phase. The nature and isotopic properties of their ices suggest inheritance from the presolar phase \citep{2019ARA&A..57..113A,2019MNRAS.490...50D}, whereas the presence of high-temperature minerals shows that comets also incorporated materials that formed close to the Sun \citep[][and references therein]{2016MNRAS.462S.323E}. 

Abundance ratios of nuclear-spin isomers for cometary molecules having identical protons have been measured in a number of comets as possible cosmogonic indicators of environmental formation conditions \citep{1987A&A...187..419M,2009ApJ...693..388K}. About two dozen measurements have been obtained for  water \citep[][and references therein]{2018AJ....156...68F} and ammonia \citep{2011ApJ...729...81S,2016MNRAS.462S.124S}. These two molecules present two nuclear-spin isomers, so-called ortho and para species. Most of the ortho-to-para ratio (OPR) determinations for water have been obtained from near-IR observations of ro-vibrational lines, using high spectral resolution.  
In the case of NH$_3$, the OPRs are estimated indirectly from observations of NH$_2$ in the optical domain.  The nuclear-spin temperatures determined from OPRs of both H$_2$O and NH$_3$ are $\sim$30 K on average \citep[e.g.][]{2018AJ....156...68F,2016MNRAS.462S.124S}. The meaning of these low spin temperatures remains debated.  
Laboratory studies of the nuclear-spin conversion of water during thermal desorption 
show fast re-equilibration in the gas phase to an OPR of 3 (which is equal to the statistical weight ratio and corresponds to a nuclear-spin temperature higher than $\sim$ 60 K) \citep{2011ApJ...738L..15H,2016Sci...351...65H,2018ApJ...857L..13H,Sliter2011}. However, the equilibrated water ortho-to-para ratio of 3, corresponding to a spin temperature higher than 60 K, is also commonly observed in comets.

In this paper, we present the measurement of the ortho-to-para ratio of water in the coma of comet 67P/Churyumov-Gerasimenko (67P) from observations performed with the high spectral resolution channel (H) of the Visible and Infrared Thermal Imaging Spectrometer (VIRTIS) on board Rosetta \citep{Coradini2007}. Because of the low activity of the comet \citep[water production rate of typically 10$^{28}$ s$^{-1}$ at perihelion; e.g.][]{2019A&A...630A..19B}, this measurement has not been possible from ground-based telescopes.  During most of the mission, VIRTIS-H acquired 2.5--2.9 $\mu$m spectra  in which ro-vibrational lines of both ortho and para H$_2$O variants are clearly present. However, the OPR retrievals are affected by optical depth effects. Hence, a careful evaluation of these effects has been performed using lines from weak vibrational bands in combination with radiative transfer calculations \citep{2016Icar..265..110D}. In Section~\ref{sec:data} the observational dataset is presented. The analysis of the spectra and the results are described in Section~\ref{sec:model-fitting}. Radiative transfer calculations are presented in Section~\ref{sec:transfer}, followed by a summary and a discussion in Section~\ref{summary}.  Results for individual VIRTIS-H data cubes are given in the appendices.

\section{VIRTIS-H observations and selected datasets}
\label{sec:data}

VIRTIS is composed of two
channels: VIRTIS-M, a spectro-imager with a visible (VIS)
(0.25--1 $\mu$m) and an infrared (IR)  (1--5 $\mu$m) channel operating at moderate
spectral resolution ($\lambda$/$\Delta \lambda$ = 70-380), and
VIRTIS-H, a cross-dispersing spectrometer providing spectra with higher spectral resolution capabilities
($\lambda$/$\Delta \lambda$ = 1300-3500) in eight
orders of diffraction covering the range 1.9--5.0 $\mu$m \citep{Drossart2000,Coradini2007}.

To determine the OPR of water, we used VIRTIS-H spectra in order numbered 4 \citep[see][for the list of orders]{2019A&A...630A..22B}, covering the wavelength range 2.432--3.077 $\mu$m in 432 spectral elements with a spectral sampling $\Delta\lambda_{\rm samp}$ = 0.0015 $\mu$m. VIRTIS-H order 4 has been found to be best suited for the OPR retrieval, as neighbouring orders either suffer from stray-light effects or have a less adequate wavelength coverage. Moreover, the cross-order absolute calibration has additional uncertainties that increase instrumental errors when several orders are in use. The nominal spectral resolving power in order 4 $\lambda$/$\Delta\lambda$ is $\sim$2200 at 2.67 $\mu$m, where $\Delta\lambda$ is the grating resolution. However, the effective spectral resolution is lower due to undersampling with respect to the grating resolution ($\Delta\lambda_{\rm samp}$ = 1.25 $\Delta\lambda$). In addition, we used calibrated data cubes where detector odd-even column readout effects  \citep{2020NatAs...4..500R} were removed by performing spectral smoothing with a boxcar average over three spectral channels, thereby reducing the spectral resolution by another factor.  From these considerations, a reduction of the spectral resolving power by a factor $\sim$ 2.7 with respect to the nominal value is expected \citep{Debout2015}. For accurate OPR determinations through spectral fitting, the good knowledge of the instrument spectral response is a key requirement.  By fitting high
signal-to-noise ratio (S/N) VIRTIS-H H$_2$O spectra of the comet with synthetic spectra (Sect.~\ref{sec:model}--~\ref{sec:fitting}) and minimising the $\chi$-square statistics, the effective spectral resolution was determined to be equal to 779 at 2.7 $\mu$m ($\Delta\lambda_{eff}$ = 0.00347 $\mu$m). 
The wavelength calibration (found to be off by up to 0.5 $\times$ $\Delta\lambda_{\rm samp}$) was also improved by adjusting the wavelengths of the spectral channels with strong water lines to expected values.

VIRTIS-H acquired data cubes of typically 3--4 h duration in various pointing modes. For the present study we considered limb observations during which the instrument stared at a fixed limb distance, with a line of sight (LOS) at an azimuth close to the comet-Sun line for most of them. The instantaneous field of view (FOV) of the VIRTIS-H instrument is 0.58 $\times$ 1.74 mrad$^2$ (58 m $\times$ 174 m for a Rosetta S/C distance to the comet of 100 km). The version of the calibration pipeline for the data is CALIBROS--1.2.  The calibration process is detailed in \citet{2015IAUGA..2255699F} and summarized in \citet{dbm2016}.

The considered data are 1) dataset S1: a set of 102 data cubes  (Table~\ref{tab:S1}) acquired from 3 June 2015 (heliocentric distance of $r_{\rm h}$ = 1.506 au) to 13 January 2016 ($r_{\rm h}$ = 2.120 au) (the perihelion of comet 67P was on 13 August 2015 at $r_{\rm h}$ = 1.242 au); the mean distances of the FOV to the comet centre ($\rho$) for these data cubes are between 1.65 and 7.13 km;  2)  dataset S2: the average of 5 data cubes obtained at far limb distances ($\rho$ = 171 to 444 km) and  mean $r_{\rm h}$ = 1.31 au; 3)  dataset S3: 28 data cubes obtained at $r_{\rm h}$= 1.561--2.730 au pre-perihelion (mean $r_{\rm h}$ = 2.292 au) that were averaged; 4)  dataset S4: average of four data cubes obtained at $r_{\rm h}$= 1.543 au post-perihelion and limb distances $\rho$ = 12.0 km ; and 5)  datasets H1 and H2: two  spectra with high S/N obtained by averaging numerous data with high-intensity H$_2$O lines acquired near perihelion.  

The resulting spectra obtained from datasets S2, S3 and S4 provide OPR measurements for a low H$_2$O column density along the LOS ($<$ 5 $\times$ 10$^{19}$ m$^{-2}$). Spectra H1 and H2 allow us to measure the OPR ratio from the faint $\nu_1$ and hot bands near 2.750 $\mu$m and longwards of 2.774 $\mu$m. Geometric informations and exposure times for datasets S2--S4 and H1--H2 are provided in Table 1.

\begin{figure*}[!htbp]
\centering
\includegraphics[width=18cm]{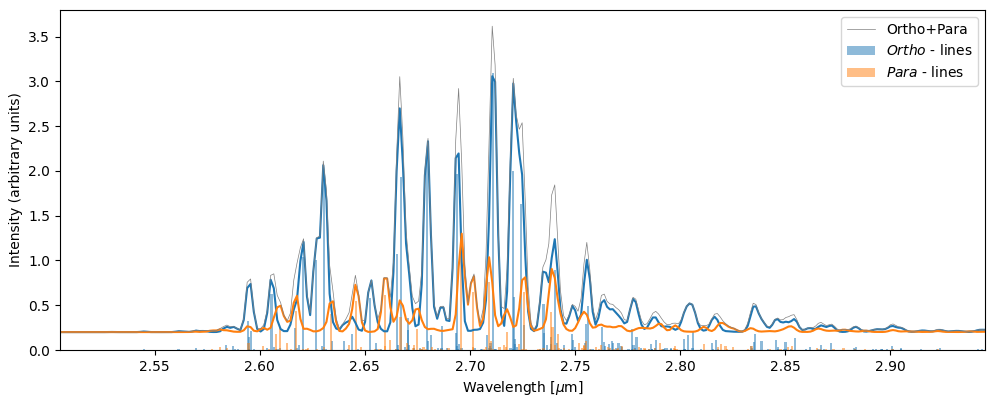}
\caption{Synthetic optically thin fluorescence H$_2$O spectrum at the effective spectral resolution (and channel spacing) of VIRTIS-H. The blue and orange lines show the ortho and para lines, respectively.   The total spectrum is plotted in grey. The ortho and para lines at infinite spectral resolution are shown at the bottom. The ortho-to-para ratio is equal to 3 and $T_{\rm rot}$ = 100 K.}\label{fig:synthetic-spectrum}
\end{figure*}

\section{Fitting method, data analysis and retrievals}
\label{sec:model-fitting}

\subsection{H$_2$O spectrum in the 2.5--3 $\mu$m range}

A typical H$_2$O spectrum acquired by VIRTIS-H near perihelion is shown in Fig.~\ref{fig:ex-spectrum}. The intense lines in the 2.55--2.76 $\mu$m range correspond to the $\nu_3$ (001-000) fundamental vibrational band. The 2.77--2.91 $\mu$m region is dominated by the weak $\nu_1$ (100-000) band, and $\nu_1+\nu_3-\nu_1$ (101-100) and $\nu_2+\nu_3-\nu_2$ (011-010) hot bands \citep{1989A&A...216..278B,Villanueva2012}. These weak bands contribute slightly to the 2.55--2.76 $\mu$m water emission. In optically thin conditions, the total H$_2$O fluorescence  emission rate (g-factor) in the 2.5--3.0 $\mu$m range at $r_{\rm h}$ = 1 au is 4.24 $\times$ 10$^{-4}$ s$^{-1}$  for a rotational temperature in the ground vibrational state $T_{\rm rot}$ = 100 K \citep[fluorescence model of][]{Villanueva2012}.

Weak emission lines from OH $v(1-0)$ prompt emission have been observed in this spectral domain in cometary spectra observed by the Infrared Space Observatory \citep{1997Sci...275.1904C,1999ESASP.427..161C}, and ground-based telescopes \citep[e.g.][]{2006ApJ...653..774B}. The strongest OH line is expected at 2.8 $\mu$m, blended with water lines \citep{1989A&A...216..278B}. Weak excess emission with respect to expected H$_2$O emission is observed at this wavelength in VIRTIS-H spectra with a high S/N (e.g. Fig.~\ref{fig:ex-spectrum}), consistent with the detection of OH prompt emission from comet 67P. Therefore, the 2.7991-2.8018 $\mu$m region of the VIRTIS-H spectra was excluded from the fit.   

\subsection{Synthetic H$_2$O spectra}
\label{sec:model}
Synthetic H$_2$O spectra were computed using the model developed by \citet{Crovisier2009}.  This model computes the full fluorescence cascade of the water molecule excited by solar radiation, describing the population of the rotational levels in the ground vibrational state by a Boltzmann distribution at $T_{\rm rot}$ (the rotational temperature $T_{\rm rot}$ is expected to be representative of the gas kinetic temperature in the inner coma of 67P). This model assumes optically thin conditions. We verified that excitation by solar radiation scattered by the nucleus and by the thermal radiation of the nucleus  can be neglected. The  model, which includes fundamental bands and hot bands, uses the comprehensive H$_2$O {\it ab initio} database of \citet{Schwenke2000} and describes solar radiation as a blackbody. The synthetic spectra closely resemble those obtained by the model of \citet{Villanueva2012}, which uses the BT2 {\it ab initio} database of \citet{Barber2006} and includes a more exact description of the solar radiation field including solar lines. However, in the 3 $\mu$m region, the excitation of the water bands is not affected by solar Fraunhofer lines (Villanueva, personal communication). When we compare spectra from the models of \citet{Crovisier2009} and \citet{Villanueva2012}, the total water emission rates in the 2.5--3 $\mu$m region differ by about 1\% ($T_{\rm rot}$ = 50 K) to 4.5\% ($T_{\rm rot}$ = 150 K). The most significant differences occur in the  the hot bands, whose fluorescence excitation and emission rates are underestimated in the model of \citet{Crovisier2009}. The total emission rates in the 2.774--2.910 $\mu$m region (referred to as the hot-band domain in Sect. ~\ref{sec:fitting}) are underestimated by 4.4\%, 8.4\%, and 9.1\% for $T_{\rm rot}$ = 50, 100, and 150 K, respectively. We checked that OPR and $T_{\rm rot}$ retrievals are similar when the  model of \citet{Villanueva2012} is used. First, by fitting synthetic fluorescence spectra from \citet{Villanueva2012} with our model as described in Sect.~\ref{sec:fitting}, the retrieved OPR and $T_{\rm rot}$ values are consistent within 1\% with the correct values. A second test consisted of modifying the fitting procedure to fit VIRTIS-H spectra with synthetic spectra from \citet{Villanueva2012} instead of spectra from \citet{Crovisier2009}: the retrievals are consistent within the uncertainties for all the output parameters of the fitting procedure (see Sect.~\ref{sec:fitting}).  

Figure~\ref{fig:synthetic-spectrum} shows a synthetic H$_2$O spectrum at the effective spectral resolution of VIRTIS-H. The ortho and para lines are plotted separately.

\subsection{Baseline subtraction and data combination}
\label{sec:baseline}

\begin{figure}[!htbp]
\centering
\includegraphics[width=9.0cm]{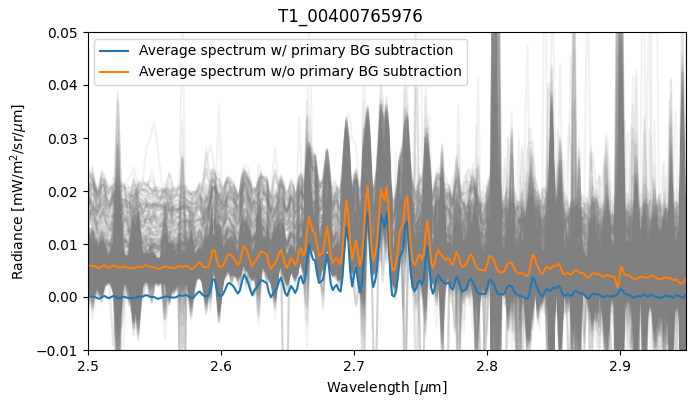}
\caption{Illustration of data combination and background subtraction. The grey spectra show the 1728 acquisitions of cube T1-00400765976 acquired on 13 September 2015 from 11h:46m:19.3s to 15h:24m:49.8s UT. The orange spectrum is the averaged spectrum, and the blue spectrum is the average spectrum obtained after subtraction of the continuum background, following the method explained in the text.  This cube, which shows extreme variations in dust continuum because it was acquired during an outburst \citep{2017MNRAS.469S.443B}, was not considered for this paper.}\label{fig:bck-removal}
\end{figure}

The water lines are observed atop an underlying continuum due to dust scattering and thermal emission \citep{2019A&A...630A..22B}. This continuum presents significant short-term variations in the time interval covered by the data cubes. In order to remove the baseline of the spectra in the best possible way, we split the observing data into cells of 32 acquisitions (the acquisition exposure time is typically 3 s). The acquisitions constituting each cell were then averaged, and the baseline was fitted by a fifth-order polynomial function considering wavelength ranges with limited H$_2$O fluorescence emission. All the baseline-subtracted cells were then averaged using a sigma clipping algorithm (with threshold of 3) to dismiss noisy spectral pixels.  Figure~\ref{fig:bck-removal} shows the individual acquisitions for a selected data cube and the final background-subtracted spectrum (in blue). The average spectrum without background subtraction is also shown (orange line).

The standard deviation for each spectral channel was evaluated using five data cubes obtained at limb distances $>$ 900 km, which present no detectable H$_2$O emission.  The statistics was performed on continuum-removed average spectra. 
 
\begin{figure*}[!htbp]
\centering
\includegraphics[width=18cm]{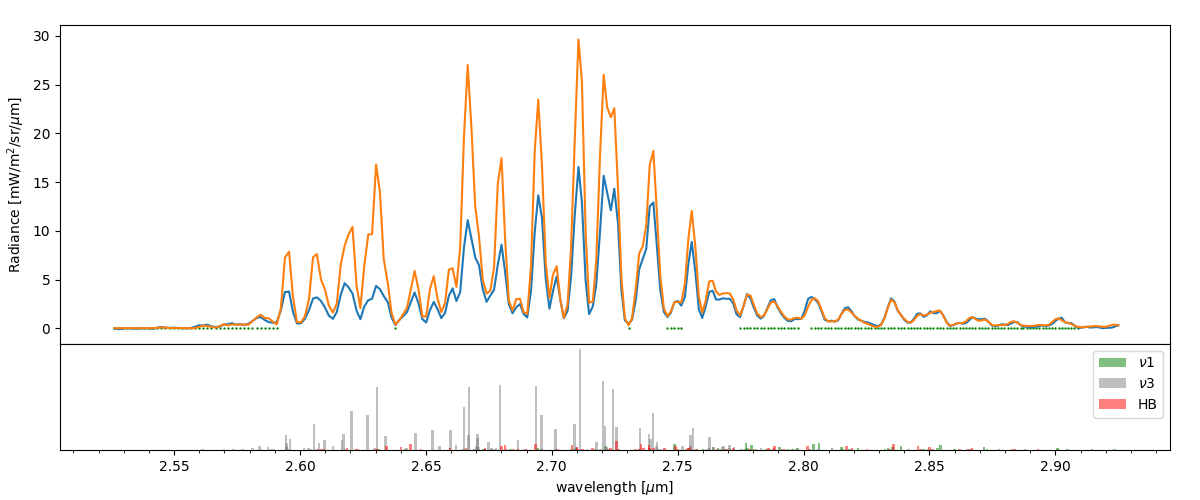}
\caption{Continuum-subtracted VIRTIS-H spectrum of dataset H1 (blue curve) superimposed on the synthetic model that best fits the HB (2.774--2.910 $\mu$m) spectral region (orange curve).  Dataset H1  is the weighted average of 22 data cubes obtained near perihelion (see Sect.~\ref{sec:H1-H2} and Table~\ref{tab:res-OPR} for geometric parameters and fitted parameters). The spectral range used to fit the HB region is indicated with a dotted green line. The bottom panel shows a synthetic optically thin spectrum at $T_{\rm rot}$ = 116 K, with different colours for the $\nu_1$ (green), $\nu_3$ (grey), and hot bands (red).}
\label{fig:ex-spectrum}
\end{figure*}

\subsection{Fitting procedure}
\label{sec:fitting}

The fitting procedure consists of fitting the  background-subtracted VIRTIS-H spectra with synthetic spectra parametrised by $T_{\rm rot}$, OPR, and a normalisation factor, and convolved to the effective  spectral resolution of VIRTIS-H (Sect.~\ref{sec:data}). The 2.590--2.760 and 2.774--2.910 $\mu$m ranges, which we refer to as the main-band (MB) and hot-band (HB) domains, respectively, were  independently fitted. The fitting procedure also corrects for  small baseline residuals. In a first step, the whole fitting procedure was applied to the HB domain (plus the 2.523--2.590 $\mu$m region with negligible water emission), using a third-order polynomial function for the baseline. Next, the MB region, without this baseline, was fitted. The derived OPR values are found to be very sensitive to baseline removal, and we optimised the removal in the best possible way.

The outputs of the fitting process are the band intensities and rotational temperatures deduced for the two domains. The OPR is a free parameter when the MB domain is fitted. It is a fixed parameter (equal to 3) when the HB domain is fitted, except for the analysis of the  H1--H2 datasets with a high-S/N (Sect.~\ref{sec:H1-H2}). The column density $N_{\rm H_2O}^{\rm MB}$ ($N_{\rm H_2O}^{\rm HB}$) is inferred from the wavelength-integrated band intensities $I^{\rm MB}$ (W m$^{-2}$ sr$^{-1}$) ($I^{\rm HB}$) using
 
\begin{equation}
I^{\rm MB} = \frac{h \nu g_{\rm thin}^{\rm MB}}{4 \pi} N_{\rm H_2O}^{\rm MB},
\label{eq:eq1}
\end{equation}

\begin{equation}
I^{\rm HB} = \frac{h \nu g_{\rm thin}^{\rm HB}}{4 \pi} N_{\rm H_2O}^{\rm HB},
\label{eq:eq2}
\end{equation}

\noindent
where $g_{\rm thin}^{\rm MB}$ and $g_{\rm thin}^{\rm HB}$ are the sum of the fluorescence emission rates (g-factors) of the individual lines in the MB and HB wavelengths domains, respectively. $g_{\rm thin}^{\rm MB}$ and $g_{\rm thin}^{\rm HB}$ were calculated from the  optically thin model of the water fluorescence (Sect.~\ref{sec:model}) using the rotational temperatures $T_{\rm rot}^{\rm MB}$ and $T_{\rm rot}^{\rm HB}$ deduced from the fitting process, and scaled according to $r_{\rm h}^{-2}$.

The presence of optical depth effects can be evaluated through the attenuation factor $f_{\rm atten}$ ($<1$ for an optically thick coma), 

\begin{equation}
f_{\rm atten}= \frac{N_{\rm H_2O}^{\rm MB}}{N_{\rm H_2O}^{\rm HB}}.
\end{equation}    

Indeed, in optically thick conditions, the effective g-factor $g_{\rm eff}$, which relates the measured band intensity $I$ (W m$^{-2}$ sr$^{-1}$) to the column density $N_{\rm H_2O}$ along the LOS, is smaller than the fluorescence g-factor $g_{\rm thin}$ \citep{2016Icar..265..110D}. Hence, the water column density derived from the MB using Eq.~\ref{eq:eq1} underestimates the actual column density by the factor $g_{\rm thin}$/$g_{\rm eff}$. On the other hand, the intensity of the $\nu_1$ and hot bands  contributing to the HB domain, which are typically one order of magnitude fainter than the main $\nu_3$ band and are therefore much less affected by optical depth effects (Sect.~ \ref{sec:model-comp-data}), provide a good estimate of the actual column density through Eq.~\ref{eq:eq2}. To first approximation, $f_{\rm atten}$ is therefore equal to the ratio $g_{\rm eff}$/$g_{\rm thin}$ of the whole $\nu_3$ band. 

\begin{figure}
\centering
\includegraphics[width=9cm]{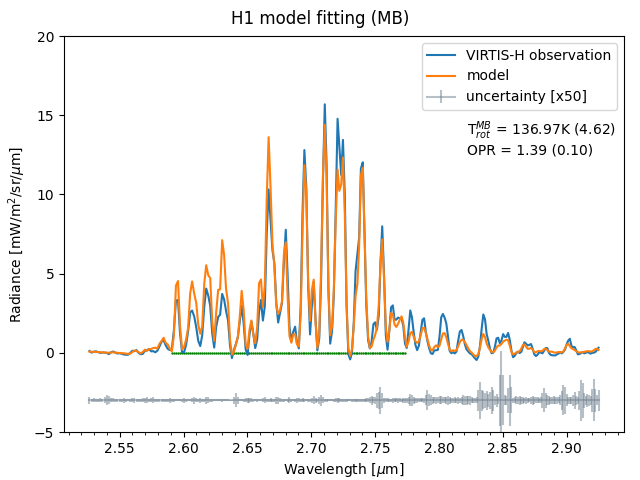}
\includegraphics[width=9cm]{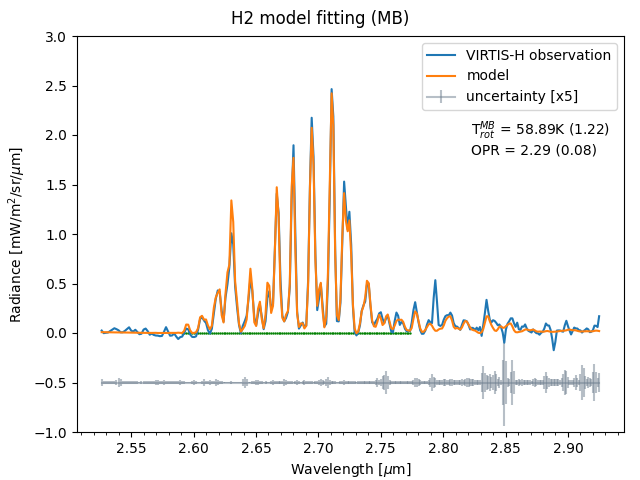}  
\caption{Continuum-subtracted VIRTIS-H spectra (blue) superimposed on the synthetic model that best fits the MB (2.590--2.760 $\mu$m) spectral region (orange). Top: H1 dataset. Bottom: H2 dataset. The standard deviation for each spectral channel is shown at the bottom of the plots (and vertically offset with respect to the measured radiances for better clarity). The MB spectral region is shown by a  horizontal green line.}
\label{fig:spectrum-MB-H1-H2}
\end{figure}

\begin{figure}
\centering
\includegraphics[width=9cm]{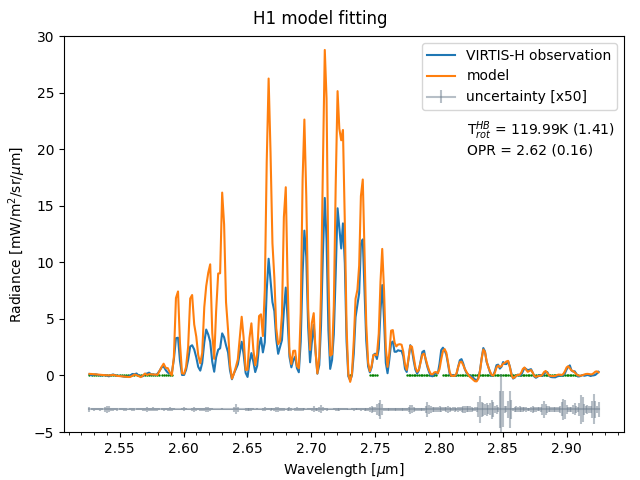}
\includegraphics[width=9cm]{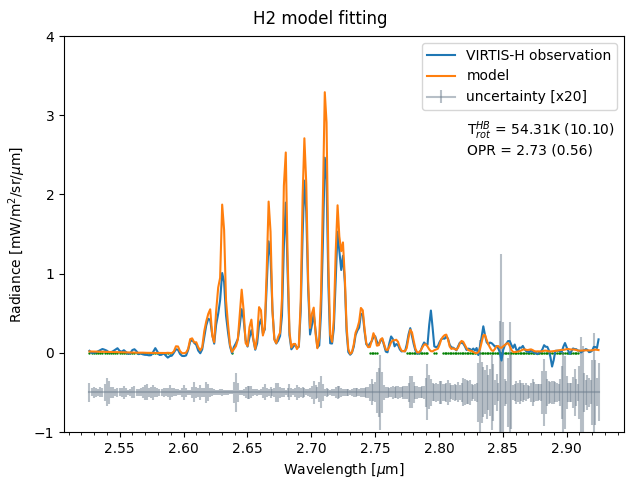}
\caption{Continuum-subtracted VIRTIS-H spectra (blue) superimposed on the synthetic model that best fits the HB (2.774–2.910 $\mu$m) spectral region (orange). Top: H1 dataset. Bottom: H2 dataset. The HB spectral region is shown by a  horizontal green line. See the caption of Fig.~\ref{fig:spectrum-MB-H1-H2} for other details. }
\label{fig:spectrum-HB-H1-H2}
\end{figure}

\begin{figure}
\centering
\includegraphics[width=9cm]{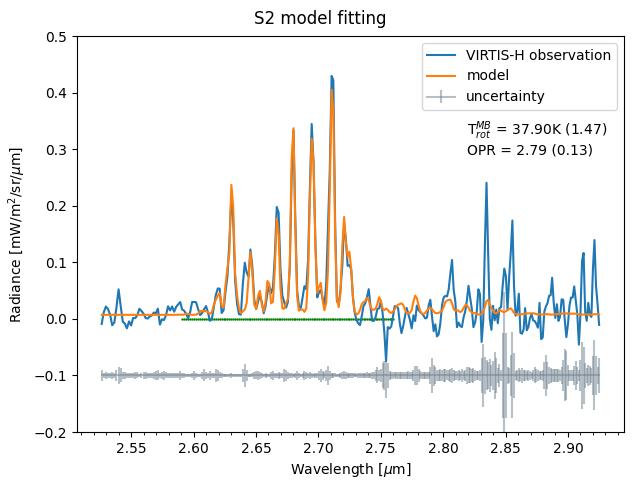}
\includegraphics[width=9cm]{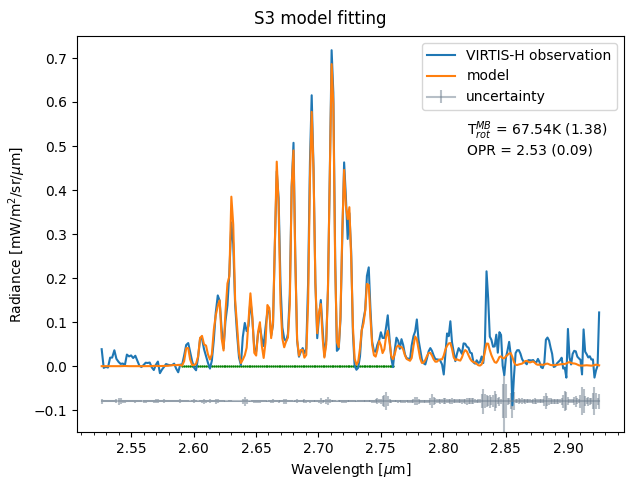}
\includegraphics[width=9cm]{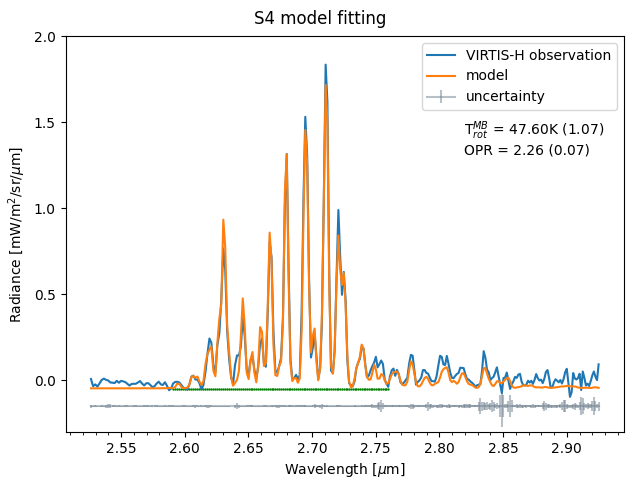}
\caption{Continuum-subtracted VIRTIS-H spectra (blue) superimposed on the synthetic model that best fits the MB (2.590--2.760 $\mu$m) spectral region (orange). Results for the S2, S3 and S4 spectra. See the caption of Fig.~\ref{fig:spectrum-MB-H1-H2} for other details.}
\label{fig:spectrum-MB-S2-S3-S4}
\end{figure}

 Spectral fits to the HB and MB wavelength ranges are shown in Figs~\ref{fig:ex-spectrum}--\ref{fig:spectrum-MB-S2-S3-S4} for datasets H1, H2, S2, S3  and S4. The inferred parameters are given in Table~\ref{tab:res-OPR}. We note that the poor fit of the MB part of datasets H1 and H2 regarding relative line intensities (Fig.~\ref{fig:spectrum-MB-H1-H2}) is characteristics of optical depth effects. A good fit is obtained when optical effects are much less significant: HB part of H1 and H2 datasets (Figs~\ref{fig:ex-spectrum} and~\ref{fig:spectrum-HB-H1-H2}) and MB part of dataset S2, S3, and S4 (Fig.~\ref{fig:spectrum-MB-S2-S3-S4}). The  attenuation of the $\nu_3$ band due to opacity is clear on  Figs~\ref{fig:ex-spectrum} and~\ref{fig:spectrum-HB-H1-H2}, especially for the H1 dataset.  Spectral fits for selected\footnote{\url{https://zenodo.org/record/6279865\#.YhjXiibjI34} for the whole set.} S1 data cubes are shown in Appendix~\ref{appendix:A}.


\subsection{Results of the fitting process}

\begin{figure}[!htbp]
\centering
\includegraphics[width=8cm]{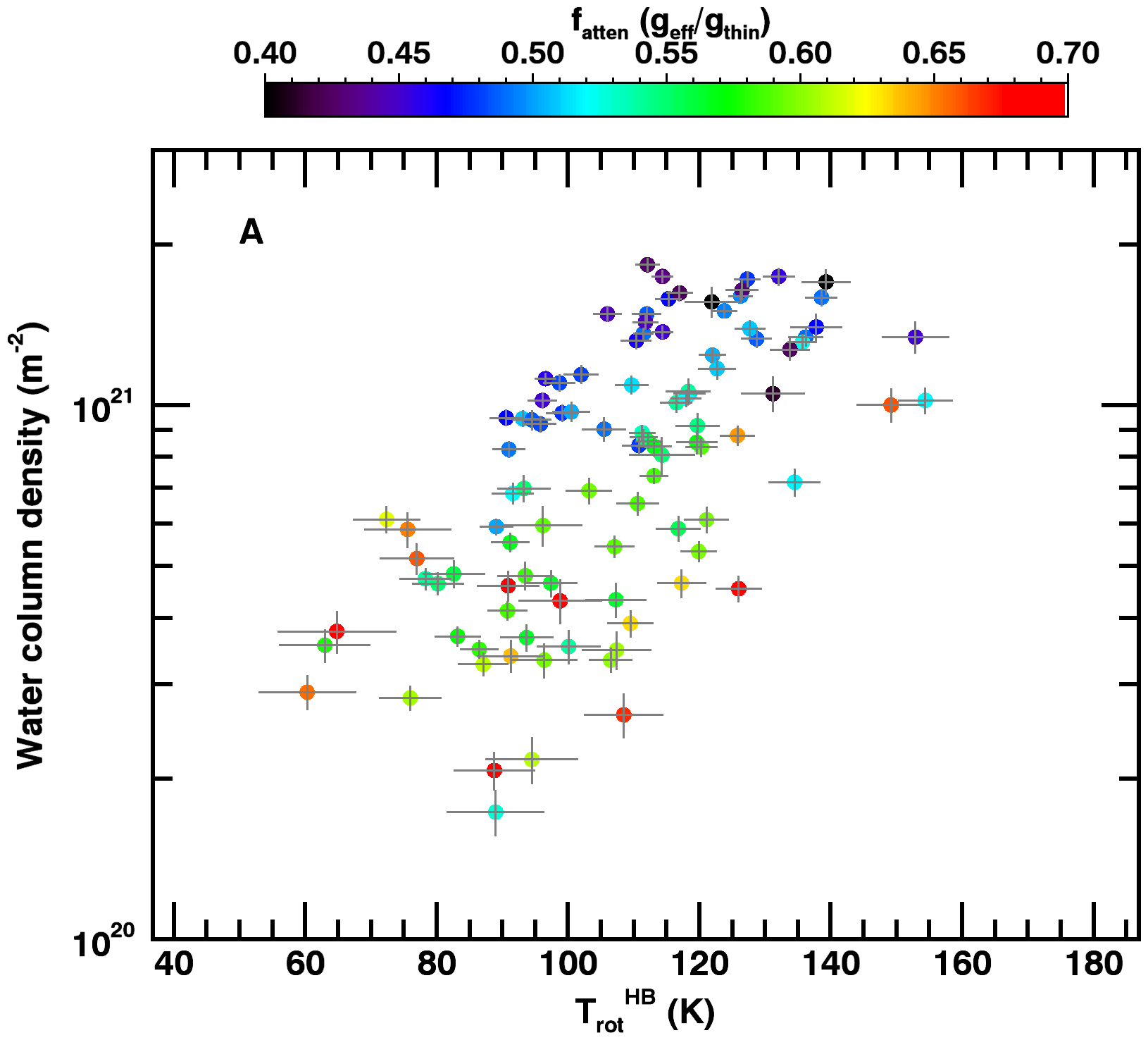}
\includegraphics[width=8cm]{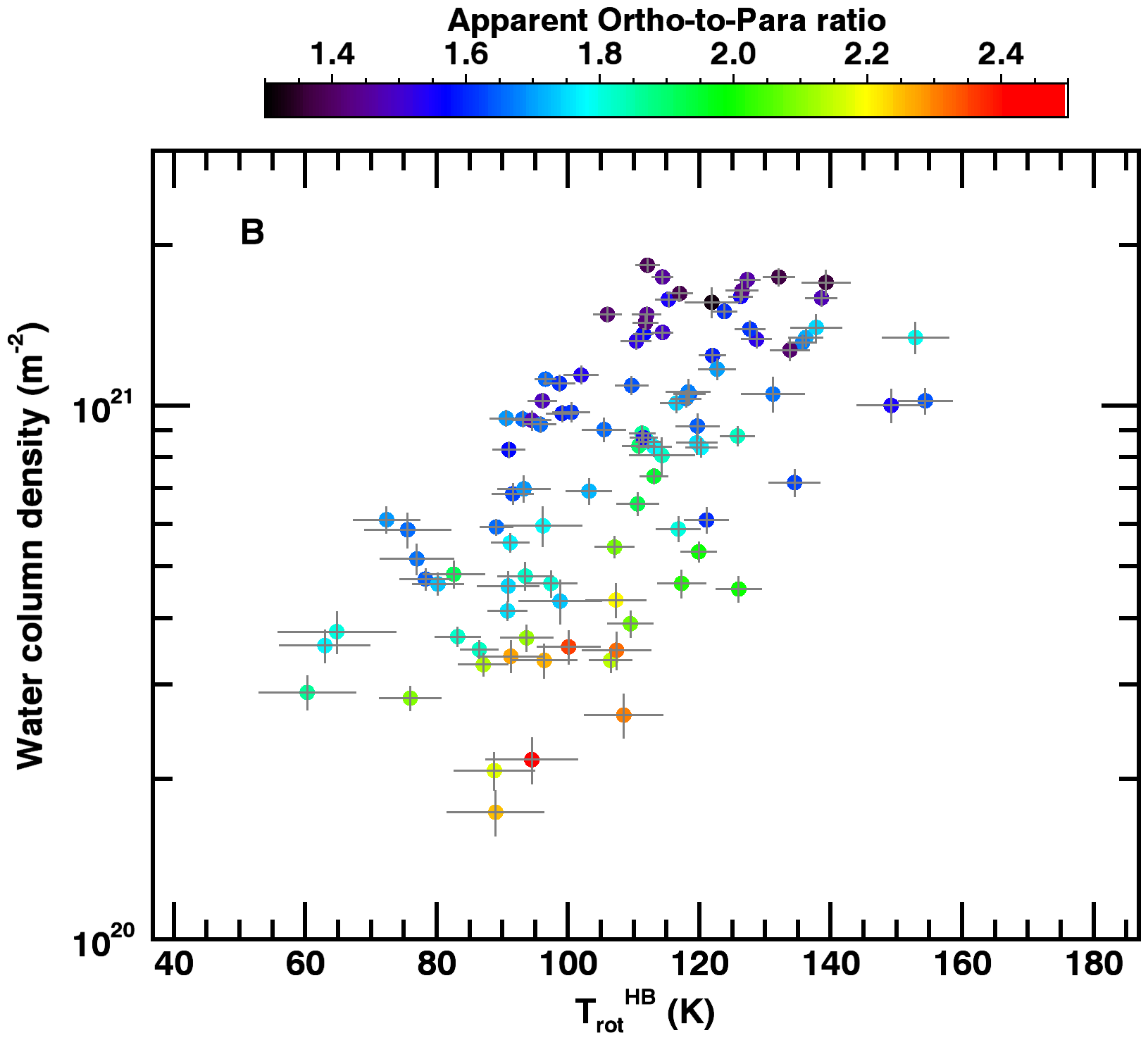}
\includegraphics[width=8cm]{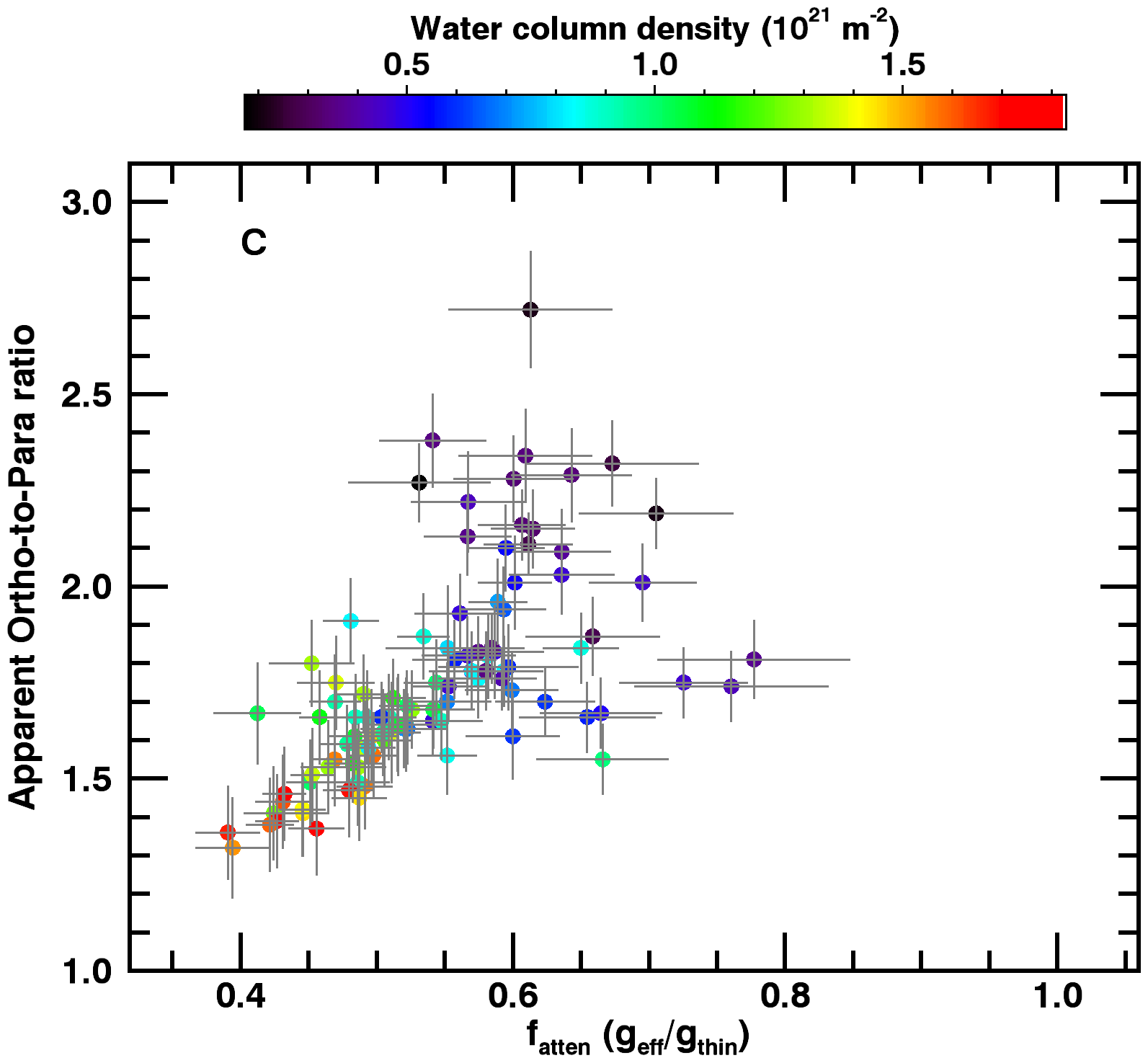}
\caption{Water column density ($N_{\rm H_2O}^{\rm HB}$), rotational temperature ($T_{\rm rot}^{\rm HB}$), apparent OPR (from MB fitting), and attenuation factor ($f_{\rm atten}$) derived from the fitting of VIRTIS-H  spectra  by optically thin fluorescence  spectra. The 102 spectra of dataset S1 are considered. Plots A--C show how these parameters vary with each other. The colour-coding is indicated at the top of each plot.}
\label{fig:fitting}
\end{figure}

\subsubsection{Dataset S1: Optically thick spectra}
\label{sec:res-S1}

Dataset S1 consists of 102 individual data cubes covering the period from 3 June 2015 ($r_{\rm h}$ = 1.506 au) to 13 January 2016 ($r_{\rm h}$ = 2.120 au).  Most spectra (92 out of 102) have been obtained at azimuth angles with respect to the comet-Sun line $PA_{\rm LOS}$ of less than 45$^{\circ}$ (Table~\ref{tab:S1}). 
This dataset has been selected for its spectra with detected signal in the HB spectral region with an S/N $>$ 10. During this time interval, the water production rate of comet 67P is seen to vary by a factor of $\sim$ 30, from typically 3 $\times$ $10^{26}$ to 8 $\times$ $10^{27}$ s$^{-1}$ according to observations carried out with the Microwave Instrument for the Rosetta Orbiter (MIRO) \citep{2019A&A...630A..19B}.  Table~\ref{tab:S1} in Appendix~\ref{appendix:B} provides geometric information for all the individual data cubes of dataset S1, and it lists the results obtained from spectral fitting.

\begin{figure}[!htbp]
\centering
\includegraphics[width=\columnwidth]{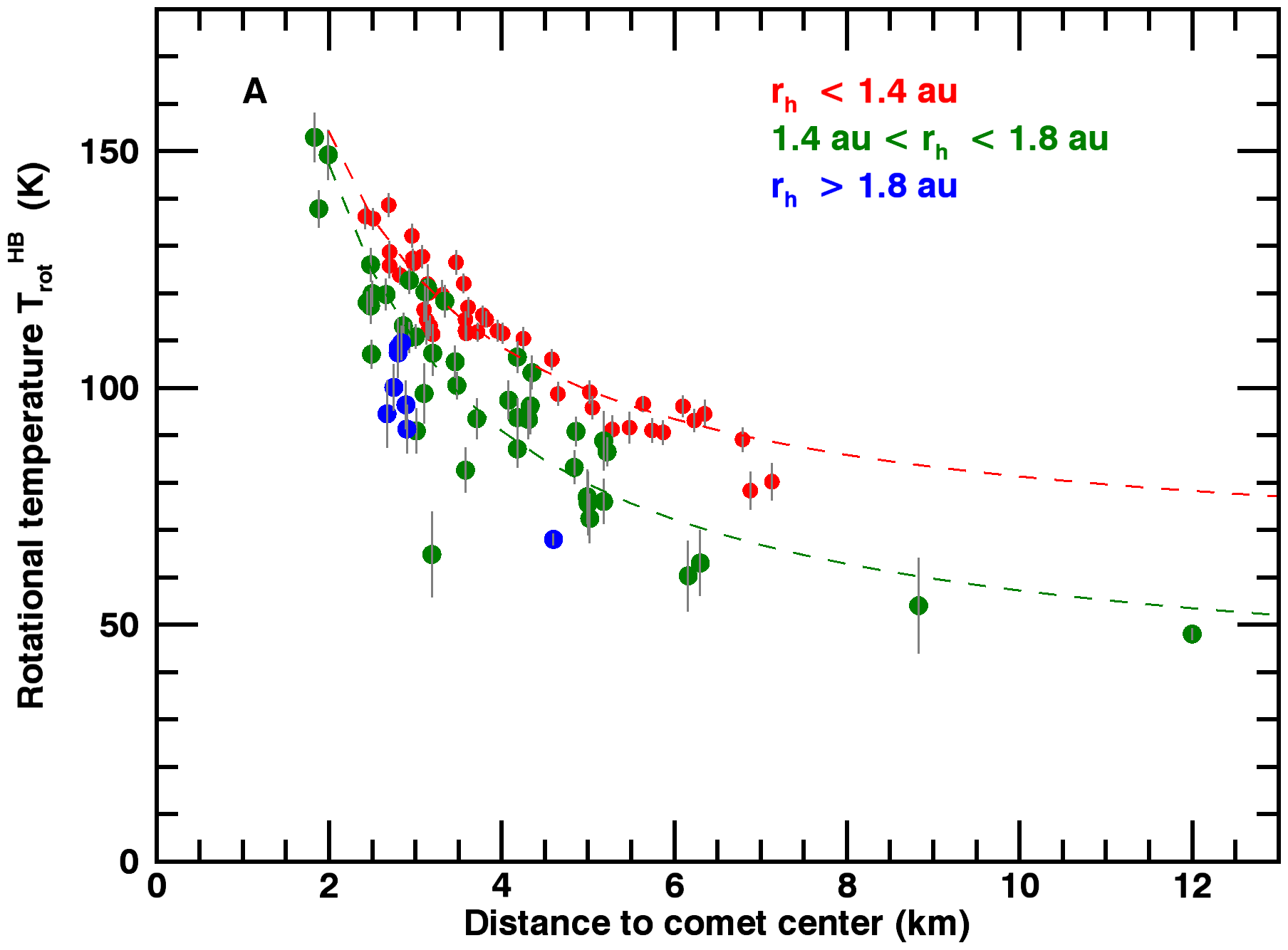}
\includegraphics[width=9cm, height=8cm]{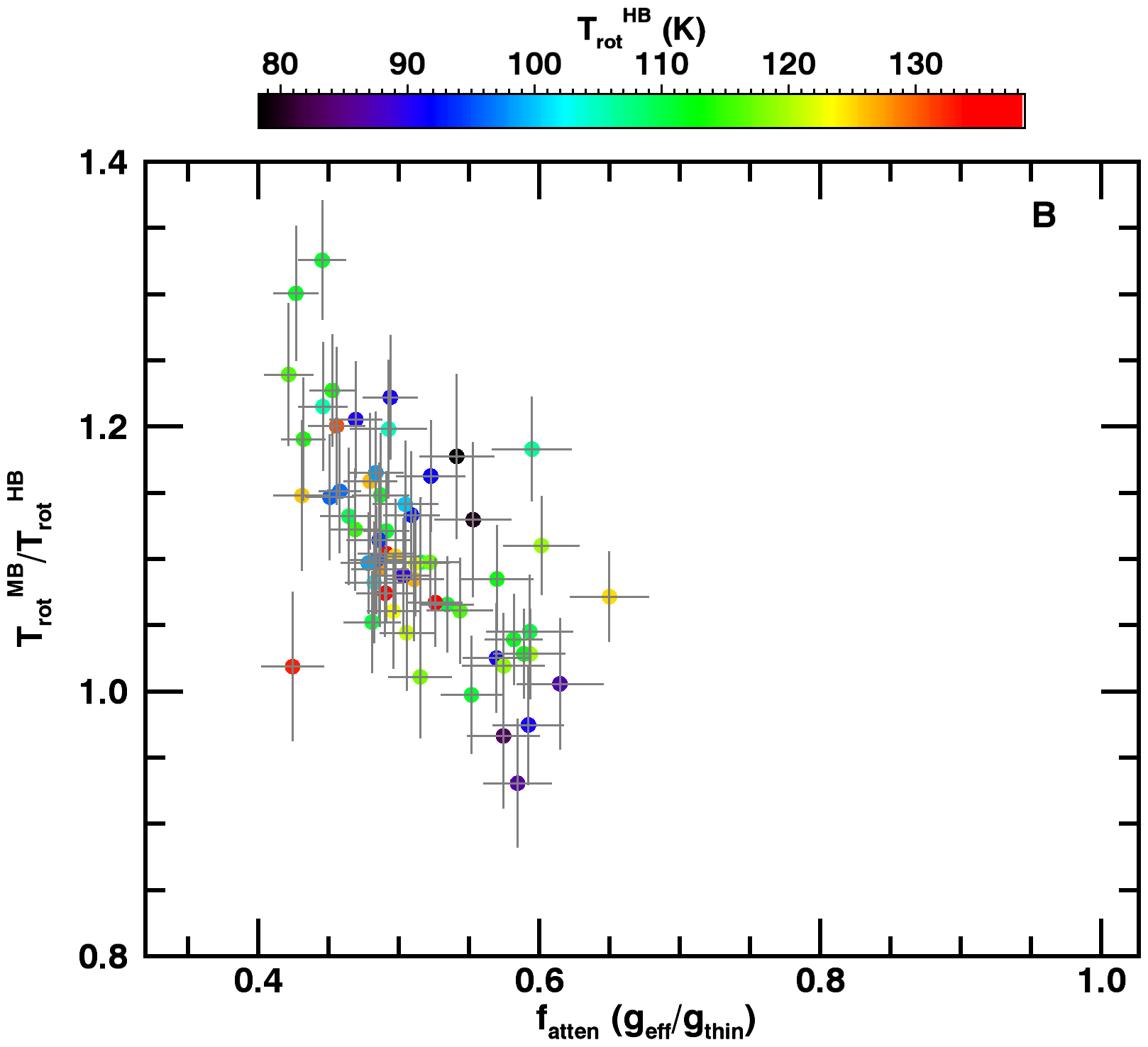}
\caption{Rotational temperatures derived from fitting the MB  (2.59--2.76 $\mu$) and HB (2.774--2.91 $\mu$m) spectral regions. A) $T_{\rm rot}^{\rm HB}$ as a function of distance to nucleus centre $\rho$, with a colour-coding according to heliocentric distance; only LOSs with a position angle with respect to the comet-Sun line $PA_{\rm LOS}$ $<$ 45$^{\circ}$  were considered; results from S3-S4 and H1-H2 spectra are included; the fitted curves are $T_{\rm rot}^{\rm HB}$ = 63 + 182/$\rho$~(K) (dashed red line for $r_{\rm h}$ $<$ 1.4 au), and $T_{\rm rot}^{\rm HB}$ = 35 + 225/$\rho$~(K) (dashed green line for 1.4 au $<$ $r_{\rm h}$ $<$ 1.8 au). B) Ratio $T_{\rm rot}^{\rm MB}$/$T_{\rm rot}^{\rm HB}$ as a function of the attenuation factor $f_{\rm atten}$  with symbol colours according to $T_{\rm rot}^{\rm HB}$. }
\label{fig:res-trot}
\end{figure}

With limb distances spanning a few hundred metres to $\sim$5 km above the nucleus surface, this dataset  
is a representative sample of the spectra obtained under highly variable LOS water column density (1.8$\times$10$^{20}$--1.9$\times$10$^{21}$ m$^{-2}$) and rotational temperature conditions (60--150 K) (Fig.~\ref{fig:fitting}A). High (small) rotational temperatures are obtained for high (low) column densities, and correspond to small (large) limb distances (Fig.~\ref{fig:res-trot}A). Rotational temperatures differ only slightly from the (number-density weighted) average of the gas temperature along the LOS  \citep{Debout2015}. 
This decrease in gas temperature with increasing altitude results from adiabatic cooling, and it was also measured by the MIRO instrument \citep{2019A&A...630A..19B}. A temperature dependence on perihelion distance is clearly observed (Fig.~\ref{fig:res-trot}A).  

Figure~\ref{fig:fitting}A shows that many water spectra are somewhat optically thick, with minimum $f_{\rm atten}$ values of $\sim$ 0.4 (equivalent optical depth $\tau$ = 0.9, with $\tau$ = --ln($f_{atten}$)). As expected for optical depth effects, there is a clear anti-correlation between $f_{\rm atten}$ and $N_{\rm H_2O}^{\rm HB}$. Because bands with lines in the HB spectral region are one order of magnitude fainter than the $\nu_3$ band, they are probably optically thin to first approximation, so that $N_{\rm H_2O}^{\rm HB}$ provides the effective water column density along the LOS. A temperature dependence of $f_{\rm atten}$ is also marginally observed (see the data points for column densities $>$ 10$^{21}$ m$^{-2}$ in Fig.~\ref{fig:fitting}A). This trend is consistent with radiative transfer calculations (Sect.~\ref{sec:transfer}). We note that $T_{\rm rot}^{\rm HB}$ should be representative of the LOS average of H$_2$O rotational temperature, whereas $T_{\rm rot}^{\rm MB}$, although much more accurately derived, is expected to be higher than $T_{\rm rot}^{\rm HB}$ when optical depth effects are significant.  Indeed, faint ro-vibrational lines excited from weakly poputated high-energy rotational levels are relatively less affected by optical depths than lines formed through low-lying levels. The relative intensities of the lines in the spectra are then  modified and mimic a warmer environnement. This trend is observed, as is clear in Fig.~\ref{fig:res-trot}B: the ratio $T_{\rm rot}^{\rm MB}$/$T_{\rm rot}^{\rm HB}$ increases with decreasing $f_{\rm atten}$, reaching $\sim$ 1.35 in the most optically thick cases corresponding to low $f_{\rm atten}$ values. For the highest values of $f_{\rm atten}$, the rotational temperature measured from the lines of the $\nu_3$ band matches the value measured from the much thinner $\nu_1$ and hot bands within 10\%.

Apparent OPRs derived from MB fitting are strongly data dependent, reaching values as low as 1.2--1.5 for LOS sampling high column densities (Fig.~\ref{fig:fitting}B). These low apparent OPRs are due to opacity effects, because lines from ortho-H$_2$O species are more efficiently weakened. OPR and $f_{\rm atten}$ follow similar trends with the H$_2$O rotational temperature and column density (Figs~\ref{fig:fitting}A, B). The decrease in optical depth effects with increasing rotational temperature is clearly visible in Fig.~\ref{fig:fitting}B: the trend for higher OPRs at higher rotational temperatures is clear over the full range of the water column density.  The strong correlation between the derived OPR and the attenuation factor is shown in Fig.~\ref{fig:fitting}C. 

At low opacity levels ($f_{\rm atten}$ $>$ 0.65), that is, at low column densities, there is significant scatter in the derived OPR. The values lie between 2.0 and 2.6. This is presumably related to spectra with low S/N.

\begin{table*}
\caption{Results from the spectral fit of datasets S2--S4 and H1--H2.}
\label{tab:res-OPR}
\begin{tabular}
{lcccccccccc} 
        \hline
       \noalign{\vskip 2mm} 
        Dataset & $t_{\rm exp}^a$ & $r_{\rm h}$ & $\rho^b$ & $N_{\rm H_2O}^{c}$ & $f_{\rm atten}$ & $T_{\rm rot}^{\rm HB}$ & $T_{\rm rot}^{\rm MB}$ & OPR$^d$ \\
                \noalign{\vskip 2mm}
             &          (hr) &        (au) &     (km) &      (m$^{-2}$) &                 &                    (K) &                    (K) &   & \\
        \hline
        \noalign{\vskip 2mm} 
        S2   & 18.0 & 1.310 &   305\phantom{00} & (6.50 $\pm$ 0.18)$\times 10^{18}$ &           - &           - &  38 $\pm$ 2 & 2.79 $\pm$ 0.13 \\
        S3   & 70.5 & 2.292 & \phantom{00}4.60 & (4.76 $\pm$ 0.08)$\times 10^{19}$ &           - &           - &  68 $\pm$ 2 & 2.53 $\pm$ 0.10 \\
        S4   & 13.6 & 1.543 & \phantom{0}12.0\phantom{0} & (4.61 $\pm$ 0.08)$\times 10^{19}$ &           - &           - &  48 $\pm$ 1 & 2.26 $\pm$ 0.07 \\
        H1   & 72.4 & 1.319 & \phantom{00}3.15 & (13.7 $\pm$ 0.34)$\times 10^{20}$ & 0.47 $\pm$ 0.01 & 120 $\pm$ 1 & 136 $\pm$ 4 & 2.62 $\pm$ 0.16 \\
        H2   & 26.9 & 1.525 & \phantom{00}8.83 & (1.23 $\pm$ 0.15)$\times 10^{20}$ & 0.62 $\pm$ 0.12 &  54 $\pm$ 10 &  59 $\pm$ 1 & 2.73 $\pm$ 0.56 \\
        \hline
        \noalign{\vskip 2mm} 
\end{tabular}

    {\raggedright
    $^a$ Total exposure time.\\
    $^b$ Mean distance of the FOV to comet centre.\\
    $^c$ For datasets S2, S3, and S4, the column density is $N_{\rm H_2O}^{\rm MB}$, derived from the intensity in the MB (2.590--2.760 $\mu$m) spectral region; for datasets H1 and H2, the column density is $N_{\rm H_2O}^{\rm HB}$, derived from the intensity in the HB (2.774--2.91 $\mu$m) spectral region. \\
        $^d$ For datasets S1, S2, and S3, the OPR is derived by fitting the MB spectral range; for datasets H1 and H2, the OPR is derived by fitting the HB range. \\
}
\end{table*}

\subsubsection{OPRs derived in low column density cases}
In the spectra of dataset S1, the H$_2$O MB is optically thick ($f_{\rm atten} < 0.65$), affecting the determination of the OPR. For these spectra, the column density is higher than $\sim$ 1.8 $\times$ 10$^{20}$ m$^{-2}$. Consequently, we attempted to derive the OPR from datasets S2, S3, and S4, for which the column density is lower by a factor 3 to 30, using the MB (the signal in the HB region is too weak to be analysed). The results are given in Table~\ref{tab:res-OPR}. The derived OPR varies between 2.3 (S4) and 2.8 (S2). The low OPR value for S4 (2.26 $\pm$ 0.07) suggests that this spectrum is more severely affected by opacity effects. The low-temperature conditions for this set ($T_{\rm rot}^{\rm MB}$ = 48 K) compared to those for set S3 
($T_{\rm rot}^{\rm MB}$ = 68 K), but similar column densities, is consistent with a higher optical depth for S4. The highest measured OPR value (2.79 $\pm$ 0.13) is for dataset S2, which samples a very low column density. The attenuation factor for these datasets is estimated in Sect.~\ref{sec:model-comp-data} using radiative transfer modelling.

\subsubsection{OPRs derived from the H$_2$O hot-band spectral region}
\label{sec:H1-H2}
Near the perihelion of 67P, lines in the 2.774--2.910 $\mu$m HB spectral region are detected with a high S/N (Fig.~\ref{fig:ex-spectrum}). These intrinsically weak lines from the $\nu_1$  and hot bands  are thinner than ro-vibrational transitions from the $\nu_3$ band, by typically one order of magnitude to first approximation  if we scale the opacity according to the relative pumping rates \citep{Crovisier2009,Villanueva2012}. Hence,  OPR values retrieved from the fit of HB spectral region are an alternative way to determine the effective H$_2$O ortho-to-para ratio for comet 67P.    
 
The H1 dataset averages 22 data cubes that satisfy the following geometric conditions: $r_{\rm h}$ < 1.4 au, 2.0 $<\ \rho\ <$ 4.0 km, and position angle $PA_{\rm LOS}$ $<$ 10$^{\circ}$ with respect to the comet-Sun line. The H2 dataset is a combination of 9 data cubes acquired from 30 May 2015 to 1 June 2015 
($r_{\rm h}$ = 1.52 au, 8.3 $<\ \rho\ <$ 9.4 km, and $PA_{\rm LOS}$ = 3$^{\circ}$. The HB domain of the H$_2$O spectrum is well detected for both sets (cf Fig.~\ref{fig:ex-spectrum} for H1). Fitted parameters are given in Table~\ref{tab:res-OPR}. The OPR determination with the best accuracy gives 2.62 $\pm$ 0.16 (H1 dataset), in agreement with retrievals obtained from S2 and S3. The attenuation factor of the HBs is estimated in Sect.~\ref{sec:model-comp-data} from radiative transfer modelling.

\section{Radiative transfer calculations and 67P water OPR}
\label{sec:transfer}

In the previous section, the VIRTIS-H data were analysed using optically thin synthetic H$_2$O fluorescence spectra. Variations in retrieved parameters OPR, $T_{\rm rot}$, and $f_{\rm atten}$ with column density and correlations between these parameters have been qualitatively explained by optical depth effects. In this section, the results of radiative transfer calculations are compared with the data and are used to derive the water OPR in the coma of 67P.  Our approach consists of computing synthetic spectra with a model considering optical depth effects both for the excitation of H$_2$O and for the received radiation. These optically thick synthetic spectra are then analysed with a fitting procedure similar to that used for VIRTIS-H data (Sect.\ref{sec:fitting}), that is, by adjusting optically thin synthetic spectra to derive the apparent OPR and rotational temperature and the attenuation factor. The approach consisting of directly fitting optically thick spectra to VIRTIS-H measured spectra was not possible, in particular, because  of the CPU time requested per model run, the variety of the coma conditions, and the large number of data cubes.   

\subsection{Radiative transfer model}
\label{sec:RTmodel}

\begin{figure} 
\centering
\includegraphics[width=9cm]{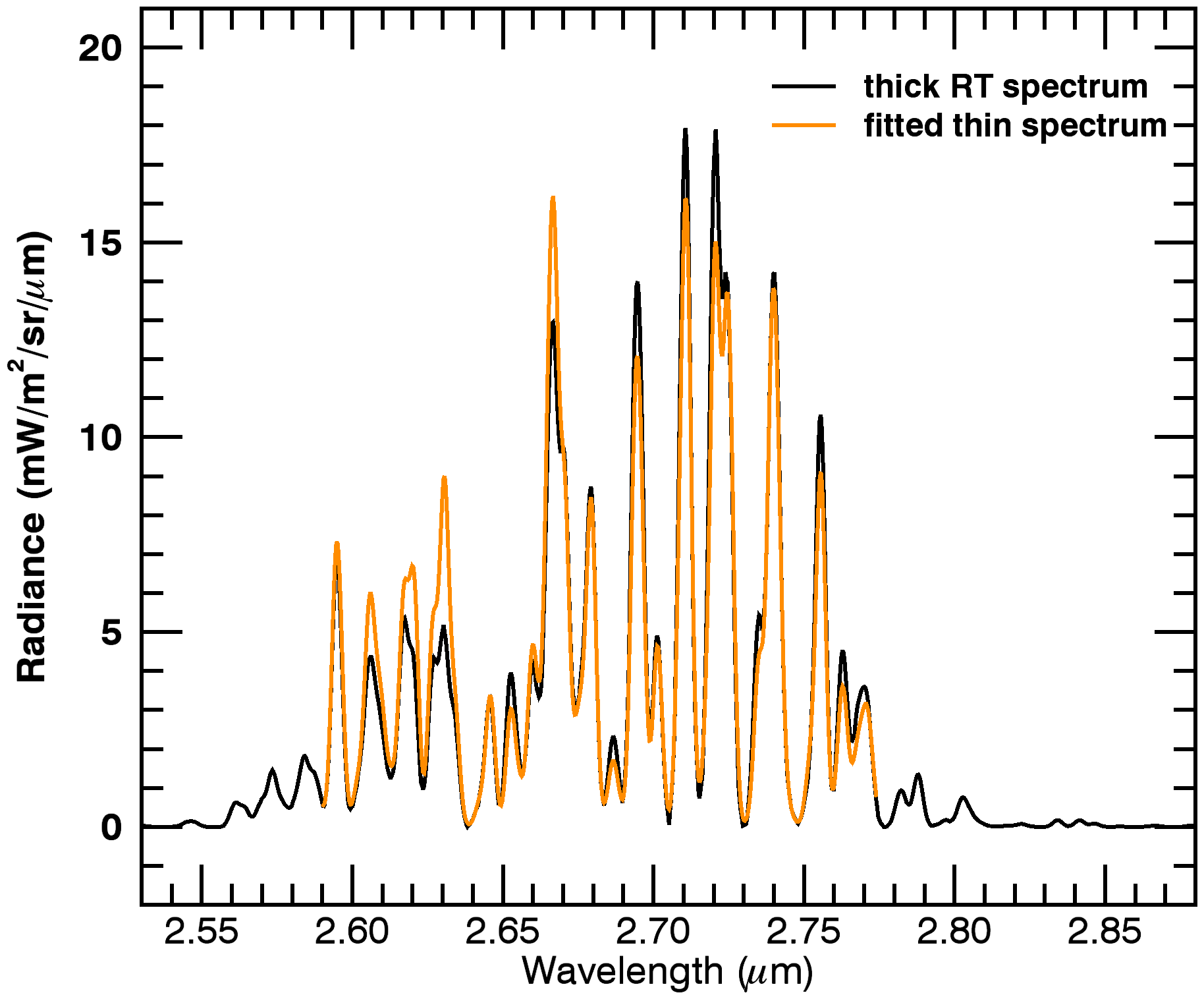}
\caption{Optically thick RT spectrum (H$_2$O $\nu_3$ band) superimposed on the fitted optically thin spectrum.
The input parameters for the calculation of the RT spectrum are OPR = 3, $N_{\rm H_2O}$ =  1.3 10$^{21}$ m$^{-2}$, an LOS mean temperature (so-called $T_{\rm LOS}^{\rm mean}$) equal to 130 K, $PA_{\rm LOS}$ = 0 $^{\circ}$, and a terminator orbit. The results of the fit are apparent OPR (i.e., OPR$^{\rm RT}$) of 1.95, attenuation factor $f_{\rm atten}^{\rm RT}$ = 0.61, and rotational temperature $T_{\rm rot}^{\rm RT}$ =  157 K. }
\label{fig:fit-RT-spectrum}
\end{figure}

\begin{figure} 
\centering
\includegraphics[width=8cm]{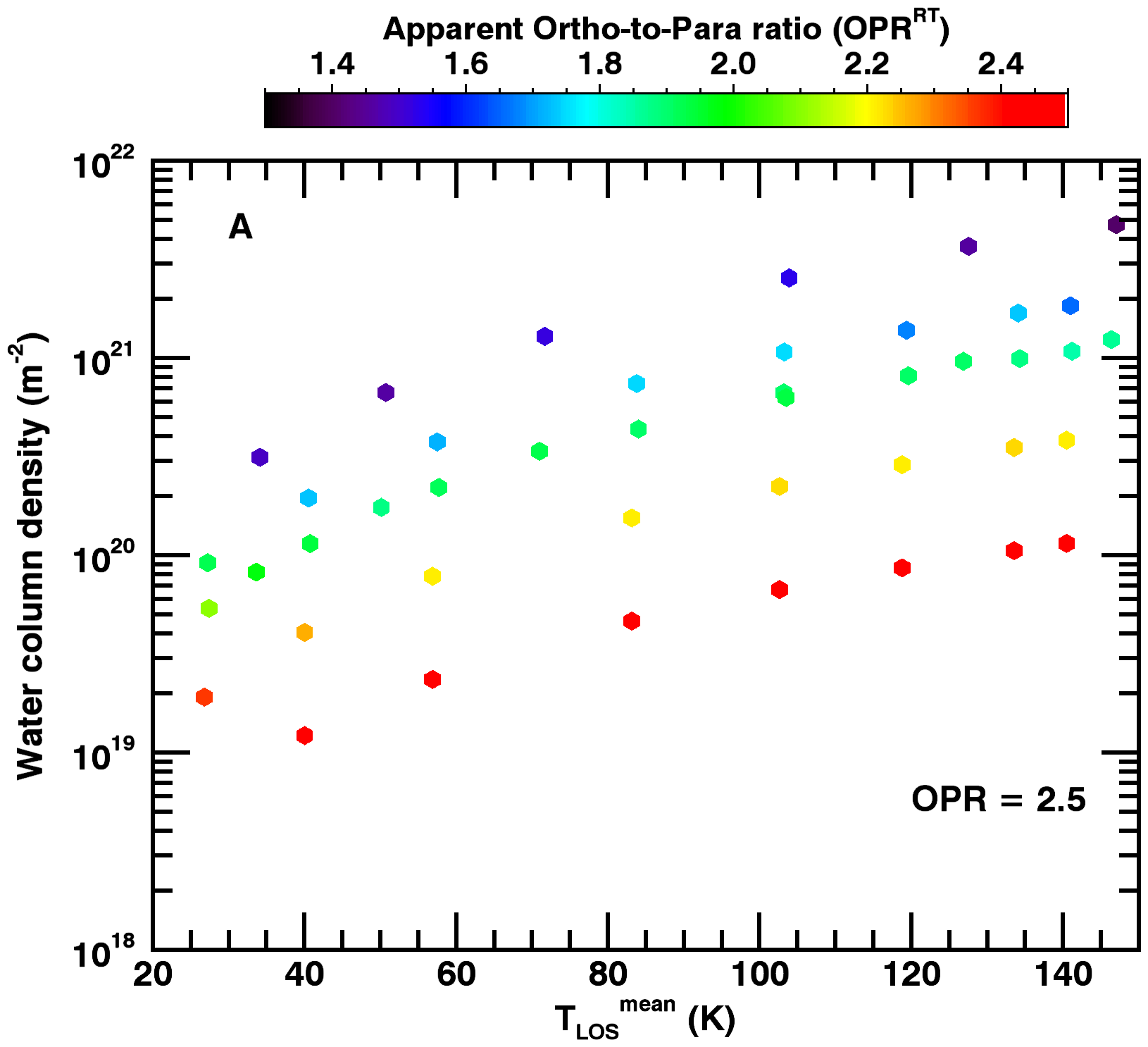}
\includegraphics[width=8cm]{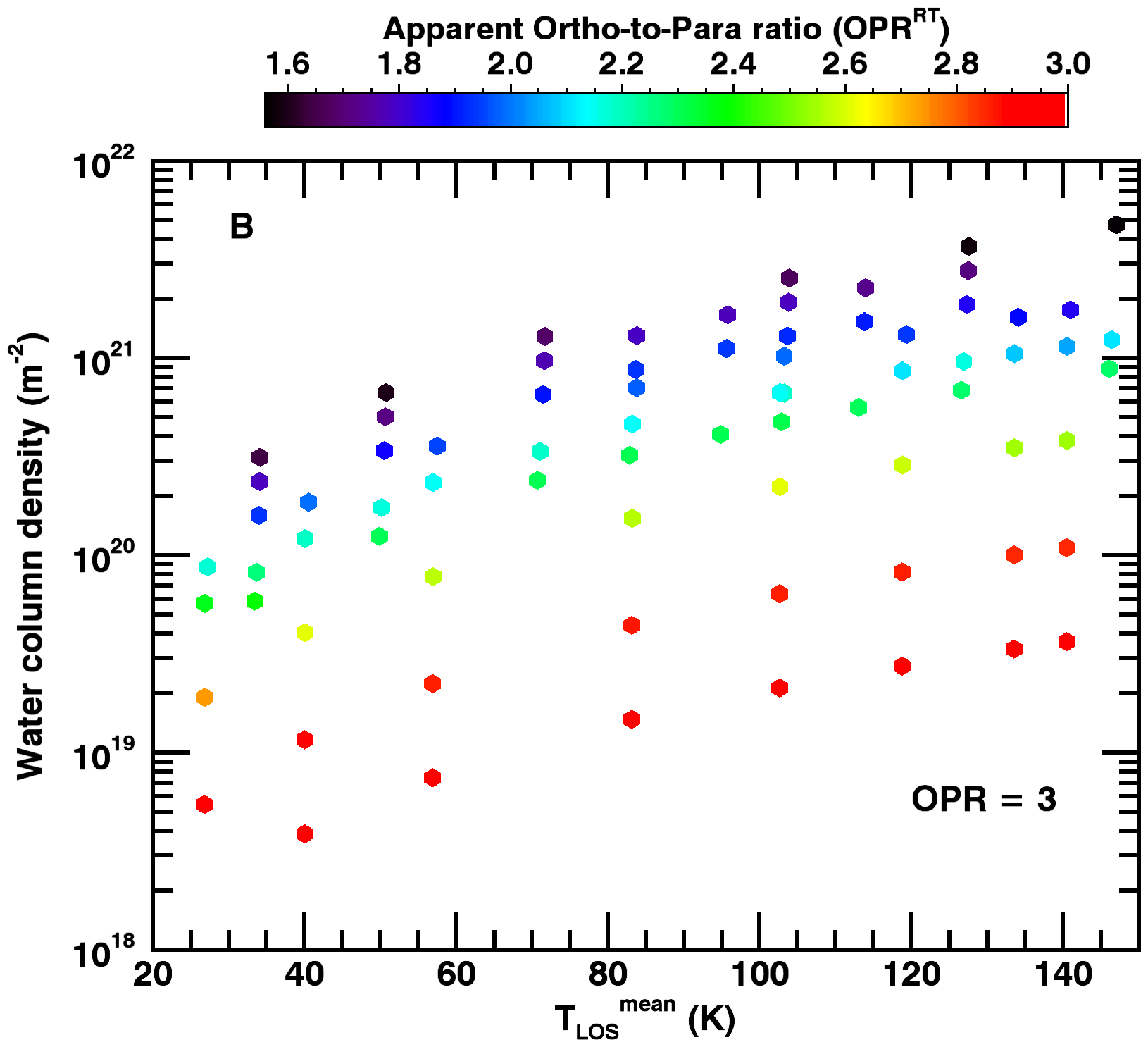}
\includegraphics[width=8cm]{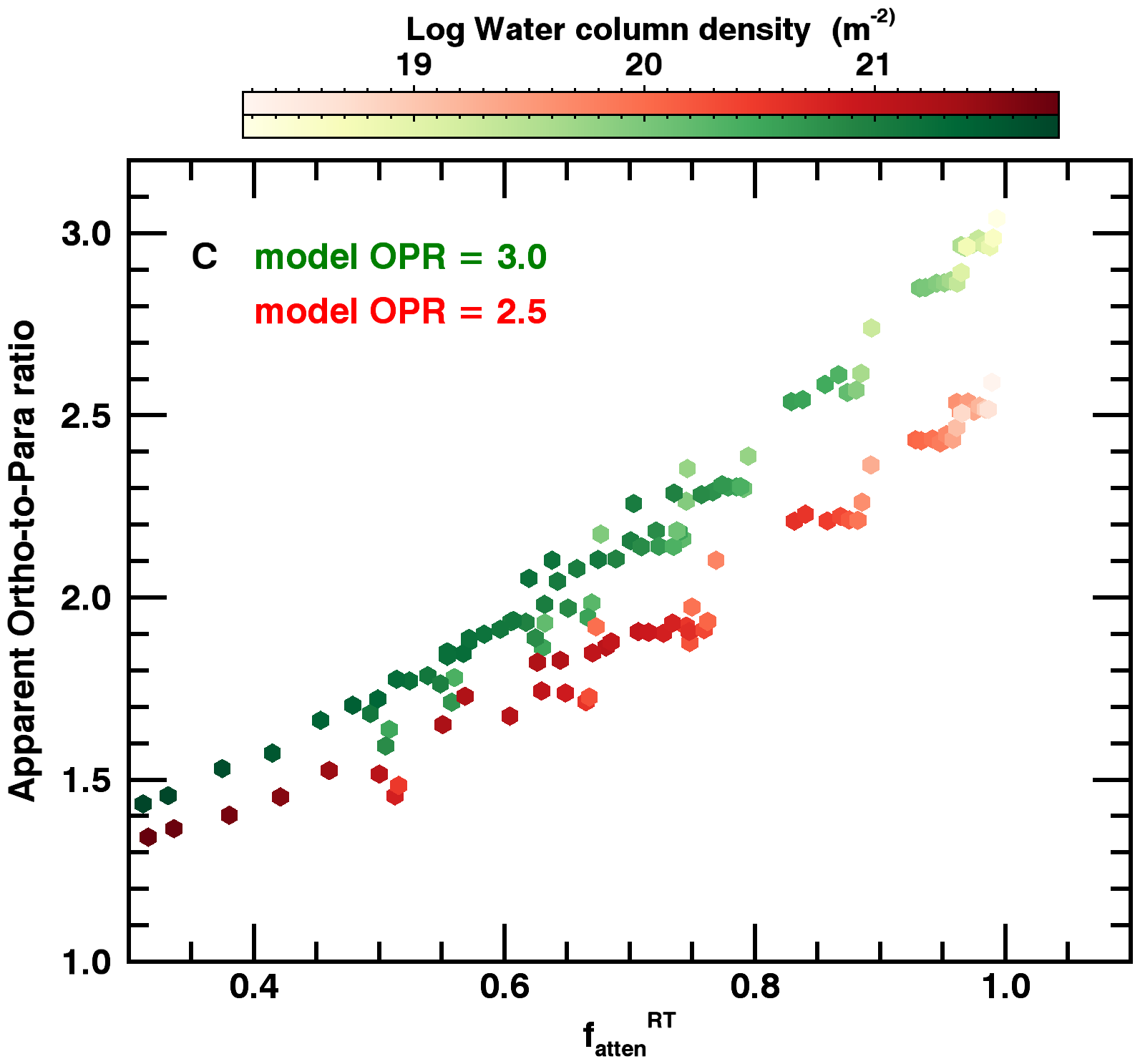}
\caption{Results from radiative transfer calculations. Apparent OPR as a function of LOS mean temperature ($T_{\rm LOS}^{\rm mean}$) and column density (A--B), and as a function of the attenuation factor (C). Plots A and B refer to calculations with an effective OPR of 2.5 and 3, respectively, with a colour-coding according to the derived apparent OPR. For C plot, the colour-coding is according to the column density (in base-10 log scale). }
\label{fig:fitting-RT}
\end{figure}

We used the dedicated radiative transfer model of \citet{2016Icar..265..110D}, which was developed for the purpose of analysing H$_2$O, CO$_2$ and CO vibrational spectra of comet 67P from VIRTIS.  Only  the $\nu_3$ band is considered for the H$_2$O model. The calculations are made for an isotropic coma, with physical parameters (number density, gas kinetic temperature and expansion velocity) derived from fluid models. As shown by \citet{2016Icar..265..110D}, optical depth effects in the solar excitation rate vary not only with the cometocentric radial distance, but also with azimuthal angle due to the  unidirectionality of the exciting source (the Sun). They are maximised along the comet-Sun direction  and minimised in perpendicular directions. The received radiation is also altered by opacity effects, which are maximum (minimum) when the phase angle (S/C-comet Sun angle) is 0$^{\circ}$ (90$^{\circ}$).  These results can be explained by the structure of the coma gas flow.  

The main input parameters of the model are the total production rate and the nucleus surface 
temperature, which determine the radial profiles of the gas kinetic temperature and velocity  (respectively decreasing and increasing with the distance to nucleus), and the gas number density \citep[see][]{2016Icar..265..110D}. We performed calculations for water production rates from 2 $\times$ 10$^{26}$ to 4 $\times$ 10$^{28}$ s$^{-1}$, heliocentric distances of 2 and 1.3 au and surface temperatures of 284 K ($r_{\rm h}$ = 2 au) and 353 K ($r_{\rm h}$ = 1.3 au). The observational geometric conditions we considered are terminator orbits and an azimuthal angle $PA_{\rm LOS}$ = 0$^{\circ}$, which correspond to most of the observing circumstances for VIRTIS-H data. Calculations were made for limb distances from 125 m to 20 km above the surface. Hence the set of computed spectra covers a wide range of column density and gas temperature conditions. We considered ortho-to-para ratios of 2.5 and 3. The spectral resolution of the synthetic spectra was 750, that is, similar to VIRTIS-H spectra.

These H$_2$O synthetic spectra (referred to as RT spectra) were analysed with a fitting procedure similar to that used for the VIRTIS-H data. However, because the RT spectra only include the $\nu_3$ band, the synthetic optically thin fluorescence spectra we used for the fitting only included this band. The fitting was performed in the MB spectral region, with $T_{\rm rot}^{\rm RT}$ and OPR$^{\rm RT}$ as output parameters.  The attenuation factor  $f_{\rm atten}^{\rm RT}$ was deduced from the ratio of the integrated intensity of the whole $\nu_3$ band to that of the  expected intensity under optically thin conditions.  We defined $T_{\rm LOS}^{\rm mean}$ as the rotational temperature obtained by not considering opacity effects in the radiative transfer model, because it is close to the LOS weighted-average of the gas temperature. $T_{\rm LOS}^{\rm mean}$ corresponds to $T_{\rm rot}^{\rm HB}$, measured on the VIRTIS data.

 Figure~\ref{fig:fit-RT-spectrum} shows an optically thick RT spectrum and the fitted optically thin spectrum. The strong similarity with the fit obtained for the MB part of H1 spectrum (Fig.~\ref{fig:spectrum-MB-H1-H2}, top) is remarkable.

\subsection{Comparison of RT model and data outputs}
\label{sec:model-comp-data}

The apparent OPR (OPR$^{\rm RT}$) is shown in Fig.~\ref{fig:fitting-RT} as a function of water column density and mean LOS temperature $T_{\rm LOS}^{\rm mean}$ for the two assumptions of the effective OPR. This figure can be compared directly to the data (Fig.~\ref{fig:fitting}), taking into account that the colour-coding is different. The same trends are observed:  the values of the apparent OPRs converge to the value of the effective OPR for water column densities significantly below 10$^{20}$ m$^{-2}$. Hence, this confirms that the determination of the effective 67P's OPR requires using results from the low-column density datasets (Table~\ref{tab:res-OPR}). The extent to which these data are affected by residual opacity effects must be examined.

\subsubsection{Direct comparison of optical depth effects in modelled and observed spectra}
\begin{figure}[!htbp]
\centering
\includegraphics[width=9.2cm]{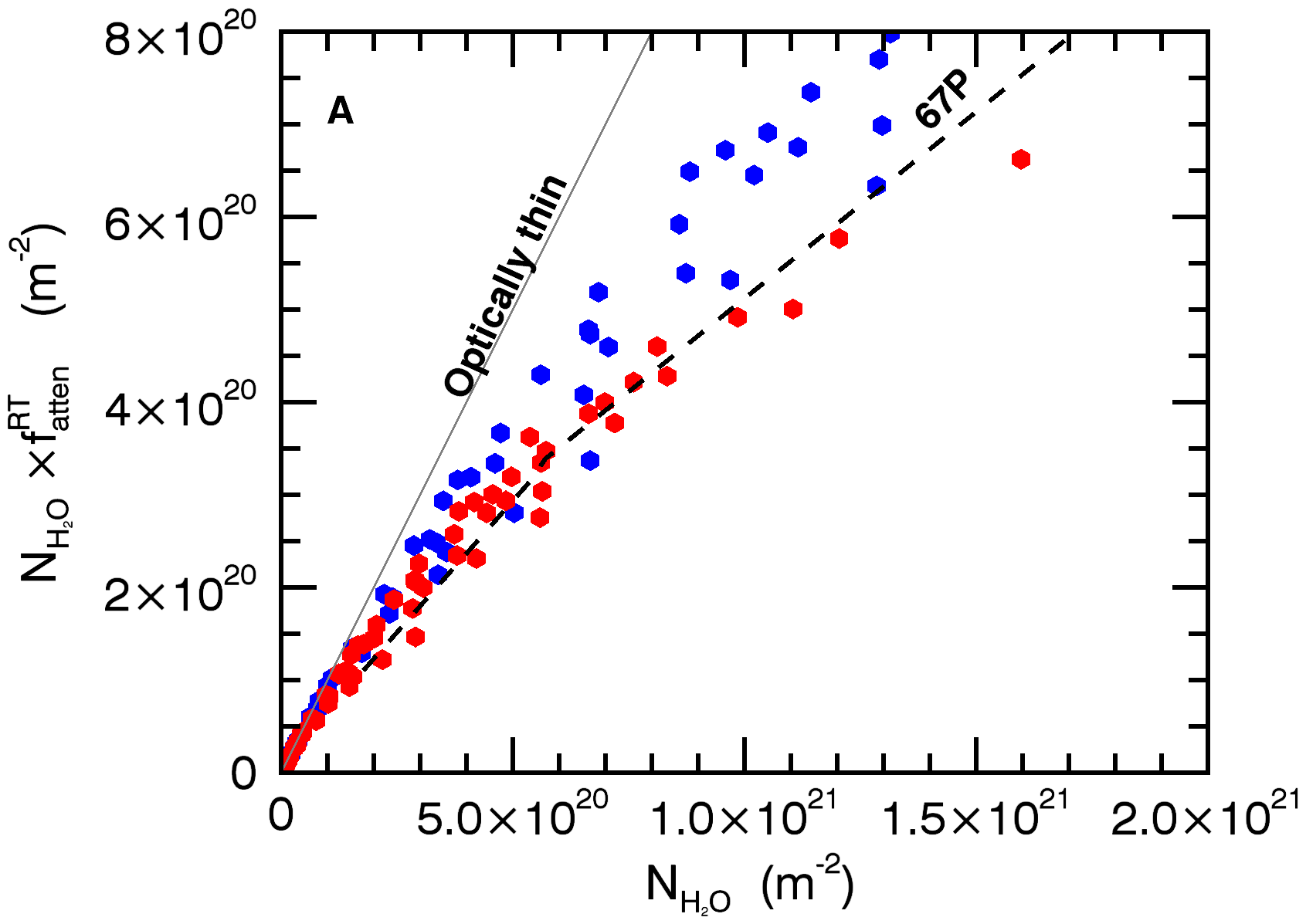}
\caption{Direct comparison of optical depth effects derived from model and observations.  The column density (blue dots) obtained by fitting the synthetic  RT spectra ($f_{\rm atten}^{\rm RT} \times N_{\rm H_2O}$) is plotted as a function of the effective column density $N_{\rm H_2O}$. The values plotted as red dots are obtained by considering that the radiative transfer model underestimates the opacity effects by a factor $f$ = 2.3 (see text). The dashed line shows the two-line fit of ($N_{\rm H_2O}^{\rm MB}$, $N_{\rm H_2O}^{\rm HB}$) values from dataset S1 (see text). The plain line shows optically thin conditions.
}
\label{fig:RT-data1}
\end{figure}

In the radiative transfer model, the assumption is made that the water coma is isotropic, which is not realistic. The water distribution observed around the nucleus of 67P looks like a broad fan approximately oriented towards the Sun \citep{2019A&A...630A..19B}. Moreover, the assumed velocity field, though physically realistic for an isotropic coma, should deviate significantly from the one in the complex coma of 67P. For these reasons, one can expect optical depth effects to be quantitatively different in the synthetic and real H$_2$O 67P's coma. Figure~\ref{fig:RT-data1} shows that this is indeed the case. We plot in this figure  (dashed black  line) a fit of $N_{\rm H_2O}^{\rm MB}$ versus $N_{\rm H_2O}^{\rm HB}$ from dataset S1. This curve is a pair of two linear regressions, one obtained by fitting data with $N_{\rm H_2O}^{\rm MB}$ $\leq 3.2 \times 10^{20}$ m$^{-2}$ (Eq.~\ref{eq:4}) and the other with $N_{\rm H_2O}^{\rm MB}$ $> 3.2 \times 10^{20}$ m$^{-2}$ (Eq.~\ref{eq:5}),

\begin{equation}
N_{\rm H_2O}^{\rm HB} = 1.761 \times N_{\rm H_2O}^{\rm MB} - 1.666 \times 10^{19},  
\label{eq:4} 
\end{equation}


\begin{equation}
N_{\rm H_2O}^{\rm HB} = 2.482 \times N_{\rm H_2O}^{\rm MB} - 2.710 \times 10^{20}. 
\label{eq:5}
\end{equation}

\noindent
The trend of $N_{\rm H_2O}^{\rm MB}$ versus $N_{\rm H_2O}^{\rm HB}$ is nicely reproduced by this two-line fit.
 
In Fig. ~\ref{fig:RT-data1}, the results from the radiative transfer model are plotted by blue dots.   Opacity effects are underestimated by the model, as is best seen for $N_{\rm H_2O} > 5.0 \times 10^{20}$ m$^{-2}$. The attenuation factor $f_{\rm atten}$ as measured from the data is about 18\% lower than for the model results. Since it is beyond the scope of this paper to change the description of the water coma in the radiative transfer model, we empirically assigned the model outputs to a column density 2.3 times lower. When this factor $f$ = 2.3 is applied, the attenuation factor given by the model  (red dots in Fig.~\ref{fig:RT-data1}) becomes consistent with the data (dashed line) for $N_{\rm H_2O} > 5.0 \times 10^{20}$ m$^{-2}$. A  slightly different factor, $f$ = 3.0, is found to  reproduce the data for $N_{\rm H_2O} < 5.0 \times 10^{20}$ m$^{-2}$ best. Interestingly the knee observed in the $N_{\rm H_2O}^{\rm MB}$ versus $N_{\rm H_2O}^{\rm HB}$ correlation is also apparent in the model results (Fig.~\ref{fig:RT-data1}). This behaviour is expected when the band opacity $\tau$, defined as $f_{\rm atten}$ = e$^{-\tau}$, varies proportionally to $N_{\rm H_2O}$. 

\subsubsection{Opacity as a function of column density and LOS temperature}

\begin{figure}[!htbp]
\centering
\includegraphics[width=9cm]{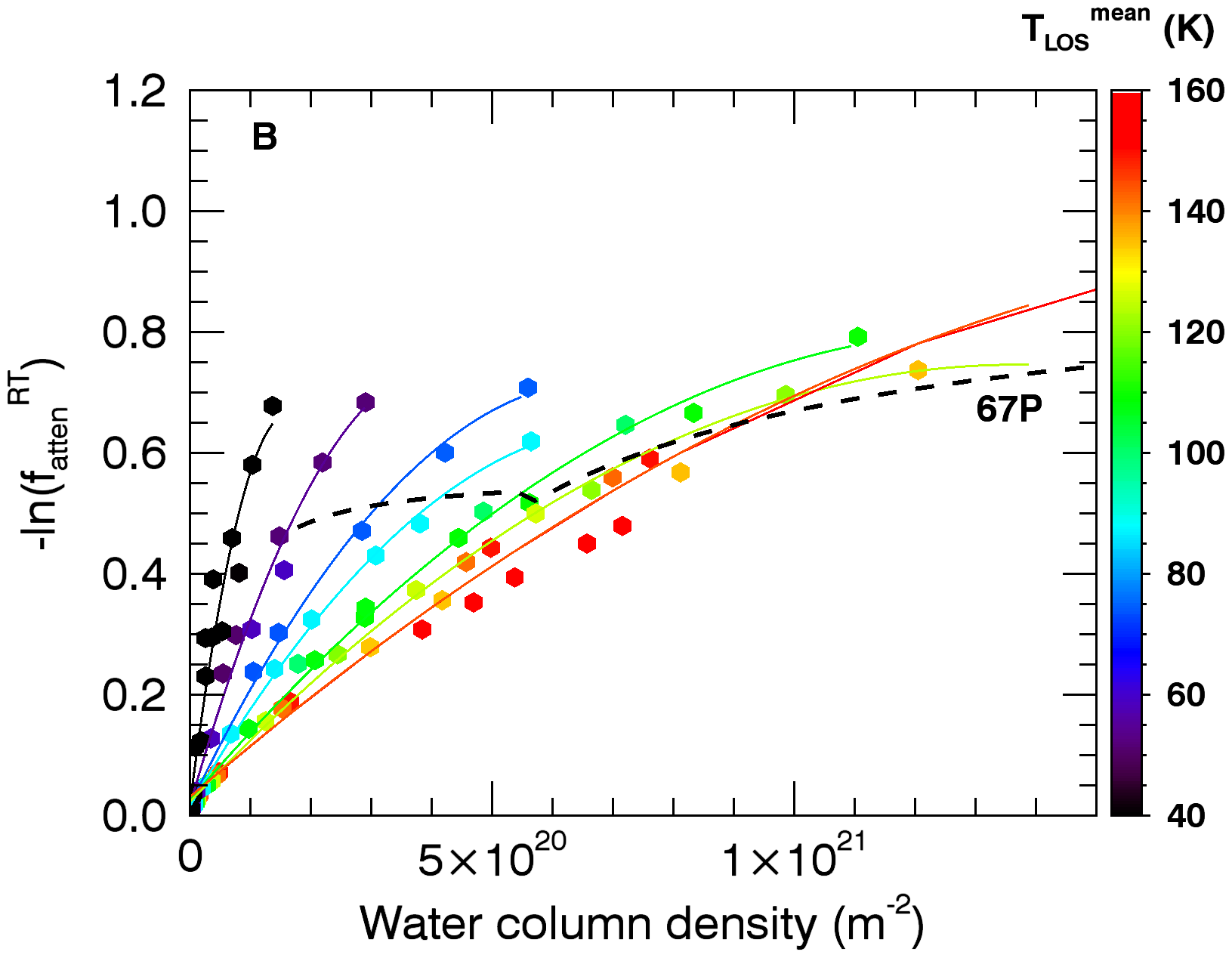}
\caption{Opacity $\tau$ = --ln($f_{\rm atten}^{\rm RT}$) as a function of column density (dots). Here, we assume that the radiative transfer model underestimates the opacity effects, and a factor of $f$ = 2.3 is applied (see text), so that the values correspond to the red dots plotted in Fig.~\ref{fig:RT-data1}. Colour-coding is according to the LOS mean temperature  $T_{\rm LOS}^{\rm mean}$, and  the colour scale is given at the right. A polynomial fitting of degree 2 is shown for various temperatures (35, 55, 70, 80, 100, 120, and 140 K). The dashed line corresponds to the two-line fit of ($N_{\rm H_2O}^{\rm MB}$, $N_{\rm H_2O}^{\rm HB}$) values from dataset S1 shown in Fig.~\ref{fig:RT-data1}.
}
\label{fig:RT-data2}
\end{figure}

Figure~\ref{fig:RT-data2} displays the opacity as a function of water column density for the model outputs  (dots), applying $f$ = 2.3 as described above, and for the data  (dashed line).
A colour-coding is used for the model outputs to show how the opacity varies with mean coma temperature along the LOS.
The opacity  increases  with decreasing LOS mean temperature. There is a non-linear correlation between opacity and water column density at fixed temperature; instead 
the trend with column density is correctly described applying a polynomial function of degree two (Fig. ~\ref{fig:RT-data2}). Comparing in this figure the VIRTIS results from dataset S1 (dashed line corresponding to the two-line fit) to the model, the agreement is satisfactory overall. The opacities derived from the data vary weakly ($\tau$ in the range 0.5--0.8) with the column density, especially for low column densities,  because the rotational temperature in the coma of 67P is lower on average  for low column densities (Fig.~\ref{fig:fitting}).

\subsection{H$_2$O OPR for 67P: Final steps}

\subsubsection{Attenuation factors of the S2--S4 main-band region}
We used the opacity variation with column density and rotational temperature shown in  Fig.~\ref{fig:RT-data2} (i.e. the polynomial functions shown by the solid lines) to estimate the attenuation factor $f_{\rm atten}$
for datasets S2--S4. Following previous discussions, calculations were also made for a correction factor $f$ = 3, as the data suggest that at low column densities the opacity  is still underestimated with a model using $f$ = 2.3. For example, for column densities $\sim 3.0 \times 10^{20}$ m$^{-2}$, the opacity derived from the data is 0.50 for a rotational temperature of 85 K, whereas the models with $f$ = 2.3 and $f$ = 3.0 give $\tau$ = 0.43 and 0.51, respectively.  However, the derived $f_{\rm atten}$ values  are not very different for the two models (see the small error bars of $f_{\rm atten}$ for S2--S4 in Fig.~\ref{fig:RT-data3}). Mean $f_{\rm atten}$ values are 0.92, 0.85, and 0.79 for S2, S3, and S4, respectively. 

\subsubsection{Attenuation factors of the H1--H2 hot-band region}
To estimate the attenuation factor for the sum of the emission lines in the HB region (about 50\% are from $\nu_1$ and 47\% are from $\nu_1+\nu_3-\nu_1$ and $\nu_2+\nu_3-\nu_2$), one has to consider that optical thickness has two sources: i) attenuation of the solar pump (ASP), and ii) absorption of emitted photons (AEP) along the ray path towards the observer. These two effects scale proportionally to $g_{lu}$ excitation and $g_{ul}$ emission  rates (g-factors), respectively, where $u$ and $l$ are the upper and lower vibrational states of the bands, respectively.  Concerning ASP, the weighted average of the $g_{lu}$ excitation rates of $\nu_1$ and  hot bands is $\sim$15 times lower than the $\nu_3$ excitation rate \citep{Villanueva2012}. As for AEP, the          
$g_{ul}$ emission rate for $\nu_1$ is $\sim$10 times lower than for $\nu_3$ \citep{Villanueva2012}. For the  hot-bands, AEP is very weak since the lower vibrational states (e.g., $\nu_1$ for the $\nu_1+\nu_3-\nu_1$ hot-band) are weakly populated \citep{2004come.book..391B}. The attenuation factor for H1 and H2 was calculated from the polynomial functions applying a multiplicative factor of 1/15 to the column density. The derived  attenuation factors ($\sim$ 0.88 for H1, $\sim 0.95$ for H2) might be slightly underestimated since AEP is weak for the hot bands.     

\subsubsection{H$_2$O OPR for 67P}
\begin{figure}[!htbp]
\centering
\includegraphics[width=9cm]{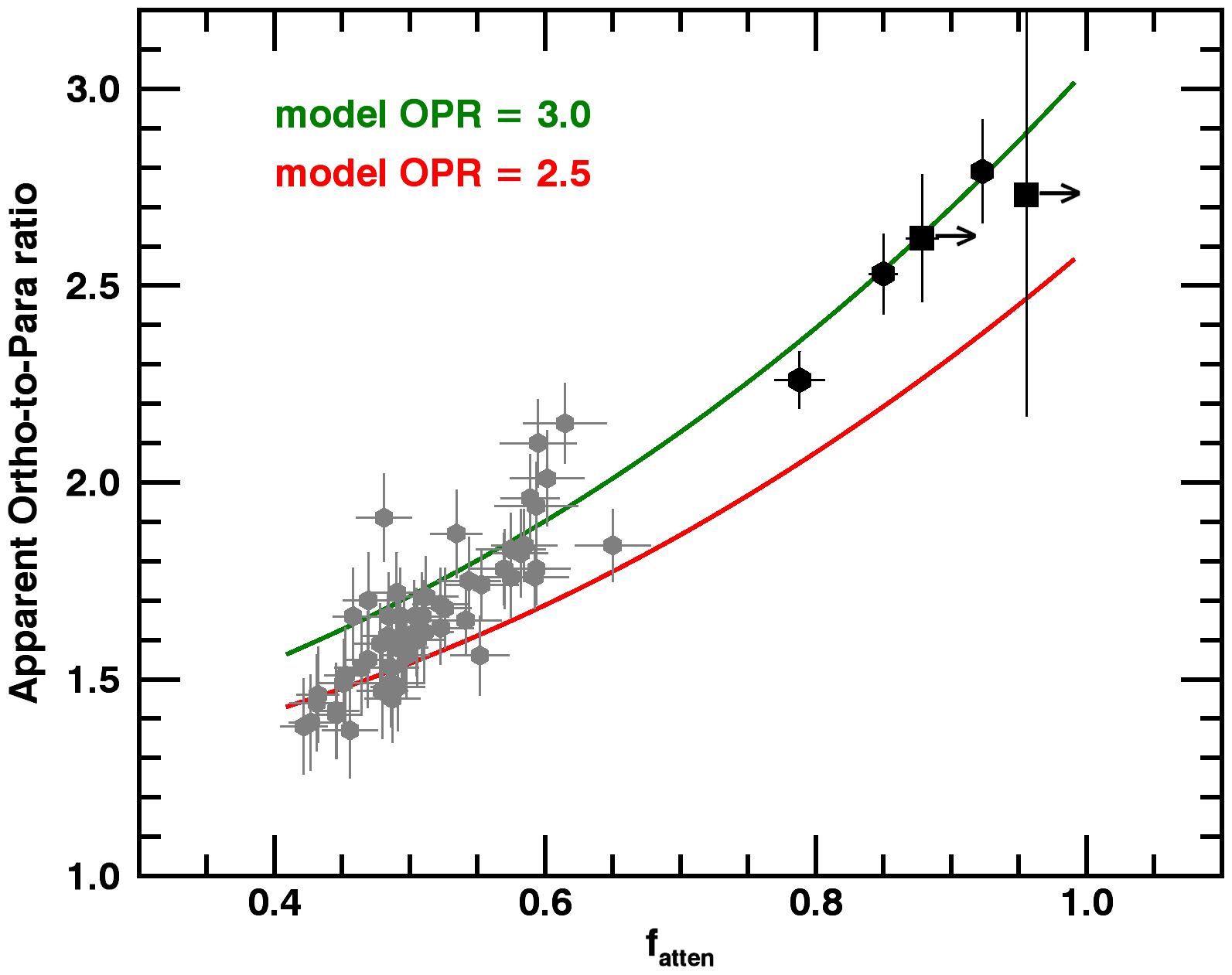}
\caption{Apparent OPR as a function of the attenuation factor $f_{\rm atten}$. Results from datasets S1 and S2--S4 are shown as grey and black dots, respectively. Results from datasets H1--H2 are shown by black squares. Only S1 data with an S/N on $I^{\rm HB}$ $>$ 20 are considered.  For S2--S4 and H1--H2, the error bars for $f_{\rm atten}$ represent the range of values obtained for $f$ = 2.3 and $f$ = 3. The arrows drawn for H1--H2 indicate a possible underestimation of $f_{\rm atten}$ for these datasets. The curves show the polynomial regressions that fit the  OPR$^{\rm RT}$ versus $f_{\rm atten}^{\rm RT}$ values shown in Fig.~\ref{fig:fitting-RT}C: red and green represent OPR = 2.5 and 3, respectively. }
\label{fig:RT-data3}
\end{figure}

The apparent OPRs as a function of $f_{\rm atten}$ for all datasets (selecting S1 data with the highest S/N in the HB domain) and for the RT models with OPR = 2.5 and 3 are shown in Fig.~\ref{fig:RT-data3}. For clarity, the results from the radiative transfer model are displayed using the polynomial regression that fits the OPR$^{\rm RT}$ versus $f_{\rm atten}^{\rm RT}$ values  shown in Fig.~\ref{fig:fitting-RT}C (instead of plotting the individual results of Fig.~\ref{fig:fitting-RT}C). This figure shows that the data points for datasets S2--S4 and H1--H2  are consistent with the RT model results obtained for OPR = 3.0. In order to estimate the OPR in the coma of 67P and its uncertainty, we searched for the polynomial regression that provides the best fit to these data points. Specifically, we assumed that the polynomial regression linking OPR$^{\rm RT}$/OPR versus $f_{\rm atten}^{\rm RT}$ that we obtain for OPR = 3 (green curve in Fig.~\ref{fig:RT-data3}), is valid for OPR values slightly different from OPR = 3. The water OPR ratio in 67P is finally estimated to 
2.94 $\pm$ 0.06. It is worth mentioning that the same result was obtained regardless of whether we considered the H1 and H2 data points because their overall weight is low.

\section{Summary and discussion}
\label{summary}

Through spectral fitting, we have investigated the H$_2$O ortho-to-para ratio in the coma of comet 67P using 2.5--3 $\mu$m  spectra obtained with the VIRTIS-H instrument on board Rosetta from 3 June 2015 to 13 January 2016. We took advantage of the detection of faint lines from the $\nu_1$, $\nu_1+\nu_3-\nu_1$ and $\nu_2+\nu_3-\nu_2$ bands beyond 2.774 $\mu$m to assess optical depth effects affecting the strong $\nu_3$ band. Most of the spectra are affected by optical thickness, and the derived apparent value for the OPR is found to decrease with increasing opacity. In the most severe conditions (H$_2$O column densities $N_{\rm H_2O}$ $>$ 10$^{21}$ m$^{-2}$), the intensity of the $\nu_3$ band is reduced by a factor of $\sim$ 2.5 and the apparent OPR is found to be as low as $\sim$ 1.3. Optical thickness also affects the ground-state rotational temperatures derived from the $\nu_3$ ro-vibrational lines: the values are found to be up to 50~K higher than those obtained from spectral fitting of the faint $\nu_1$ and hot-bands. Observed trends with H$_2$O column density and mean gas coma temperature along the LOS are consistent  with radiative transfer calculations, although quantitative differences persist, presumably related to the simplistic description of the coma of 67P that was used for these calculations.

Three spectra (S2, S3, and S4), corresponding to the average of data cubes fulfilling low-opacity conditions ($N_{\rm H_2O}$ $<$ 5 $\times$ 10$^{19}$ m$^{-2}$), were used to infer the effective water OPR. In addition, we combined numerous spectra with high S/N (dataset H1), to measure the OPR from lines of the $\nu_1$ and  hot-bands. According to our radiative transfer calculations, the range of OPR values derived from these datasets (2.26 to 2.79) corresponds to small to moderate optical depth effects. The H$_2$O OPR in comet 67P was finally estimated to be 2.94 $\pm$ 0.06, which is consistent with the statistical value of 3.

The H$_2$O OPR has been measured in numerous comets \citep[see][for the most recent compilations]{2018AJ....156...68F, 2019MNRAS.487.3392F}. For several comets, values are near 2.5, corresponding to a spin temperature of $\sim$30 K. However, in other comets, estimates are consistent with the statistical value.  \citet{2019MNRAS.487.3392F} found that whereas the mean value in a sample of 19 comets is 2.62$\pm$0.03, the median value is 2.86, which is not far from the statistical value. Unequilibrated and equilibrated values are found for both Oort cloud and Jupiter-family comets.
Most OPR determinations are from ground-based observations of ro-vibrational lines from H$_2$O hot bands near 2.9 $\mu$m. These determinations should not be significantly affected by optical depth effects, but a quantitative investigation remains to be performed.  Interestingly, spin temperatures measured for NH$_3$ are also clustered near 30~K, and a good correlation between spin temperatures of H$_2$O and NH$_3$ is observed 
in the range of spin temperatures from 24 K to $>$ 40 K \citep{2016MNRAS.462S.124S}.

Whereas in the past, OPRs in comets were believed to be of cosmogonic significance, that is, related to the formation history of cometary ices \citep[e.g.][]{2004come.book..391B}, their real meaning is now unclear.
  \citet{Sliter2011} measured the OPR of H$_2$O monomers thermally desorbed from ice. They isolated H$_2$O in an Ar matrix at 4 K and left the condensed sample for about a day to obtain almost purely para isomers of H$_2$O. Subsequently, the Ar matrix was sublimated by rapid heating, and the IR spectrum of the water vapour sublimated from the ice at 250 K was recorded. It was found that the relative intensities of the rovibrational lines were equivalent to those of a H$_2$O vapour at room temperature. Thus, it was concluded that the conversion from para to ortho in ice is considerably fast. \citet{2016Sci...351...65H,2018ApJ...857L..13H} prepared H$_2$O ice at 10 K in two different ways: vapour deposition of H$_2$O and hydrogenation of O$_2$. In both cases, the OPR of H$_2$O photo or thermally desorbed from ice at 10 K and 150 K, respectively, was found to be the high-temperature-limit value of 3. Theoretical studies suggest that nuclear-spin conversion results from intermolecular proton–proton magnetic dipolar interactions and occurs on very short time scales \citep[$\sim$ 10$^{-5}$--10$^{-4}$ s;][]{Bunt2008} in ice. In other words, the OPR of H$_2$O desorbed from cometary nuclei should have the statistical value of 3, regardless of the past formation process of the ice.

As discussed by \citet{2016MNRAS.462S.124S}, some processes in the coma might change the OPR after the molecules sublimated from the cometary surface. Nuclear-spin conversions by radiative transitions are strictly forbidden, but 
might occur by interactions with protonated ions such as H$_3$O$^+$, water clusters, ice grains, and paramagnetic materials (e.g. O$_2$ molecules and some minerals in dust particles) in the near nucleus collisional region \citep{2000prpl.conf.1159I,Hama2013}. \citet{2007ApJ...661L..97B,2008Icar..196..241B} measured the water OPR
as a function of cometocentric distance in comets 73P-B/Schwassmann-Wachmann 3 and C/2004 Q2 (Machholz) and did not observe any change in the ranges 5--30 km (73P) and 50-800 km (C/2004 Q2). For these two comets, the OPR is close to the statistical value. Most of the VIRTIS-H limb data have been   acquired within $\rho$ = 10 km from comet centre. OPR variations that could be due to nucleus-spin conversion, if any, cannot be brought out from these close-nucleus data, which are affected by optical depth effects. However, the sensitive measurement obtained at $\rho$ = 300 km  near perihelion (dataset S2, at $r_{\rm h}$ = 1.31 au) is consistent with the other low-opacity data acquired much closer to the nucleus (Table~\ref{tab:res-OPR}), and does not support significant ortho-to-para conversion in the inner coma of comet 67P.  Some comets present unequilibrated OPR values at close cometocentric distances. For example, a value of 2.59 $\pm$ 0.13 was measured for comet 103P/Hartley 2 from near-IR data sampling projected distances $<$ 66 km from the nucleus  \citep{2013Icar..222..740B}. Hence, nuclear-spin conversion might occur in the very inner coma of some comets (and possibly on the surface by interaction with paramagnetic impurities) by a process that remains to be identified. This is not observed for comet 67P, however.

\begin{acknowledgements}
The authors would like to thank the following institutions and
agencies,  which supported this work: Italian Space Agency (ASI -
Italy), Centre National d'Etudes Spatiales (CNES -- France),
Deutsches Zentrum f\"{u}r Luft- und Raumfahrt (DLR -- Germany),
National Aeronautic and Space Administration (NASA -- USA). VIRTIS
was built by a consortium from Italy, France and Germany, under
the scientific responsibility of the Istituto di Astrofisica e
Planetologia Spaziali of INAF, Rome (IT), which lead also the
scientific operations.  The VIRTIS instrument development for ESA
has been funded and managed by ASI, with contributions from
Observatoire de Meudon financed by CNES and from DLR. The
instrument industrial prime contractor was former Officine
Galileo, now Leonardo company in Campi Bisenzio,
Florence, IT. The authors wish to thank the Rosetta Science Ground
Segment and the Rosetta Mission Operations Centre for their
fantastic support throughout the early phases of the mission. The
VIRTIS calibrated data are available through the ESA's
Planetary Science Archive (PSA) Web site and NASA Planetary Data System (PDS). Y.-C. Cheng aknowledges funding from IRIS OCAV (Universit\'e PSL). The work of M. Roos was supported by the program DIM-ACAV of R\'egion Ile de France. With fond memories of
Angioletta Coradini, conceiver of the VIRTIS instrument, our
leader and friend, and to Michel Combes, one of the first architects of the  VIRTIS-H channel.
\end{acknowledgements}

%
%

\clearpage
\begin{appendix}

\FloatBarrier

\onecolumn
\section{Example of spectral fits to S1 cubes.}
\label{appendix:A}

\FloatBarrier

\begin{figure*}[hbt!]
\centering
\begin{minipage}{9cm}
\includegraphics[width=9cm]{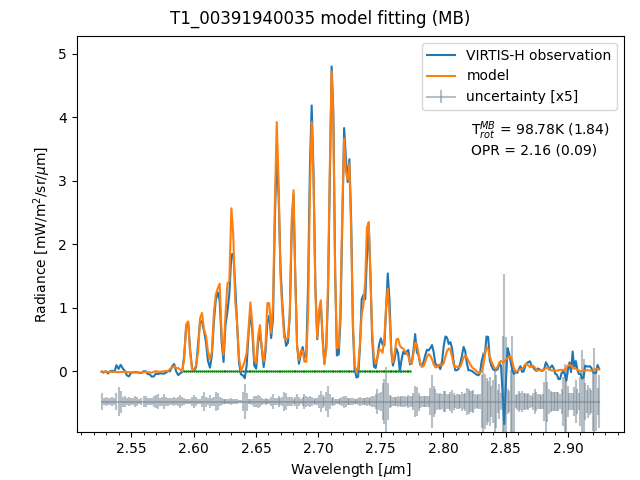}
\end{minipage}\hfill
\begin{minipage}{9cm}
\includegraphics[width=9cm]{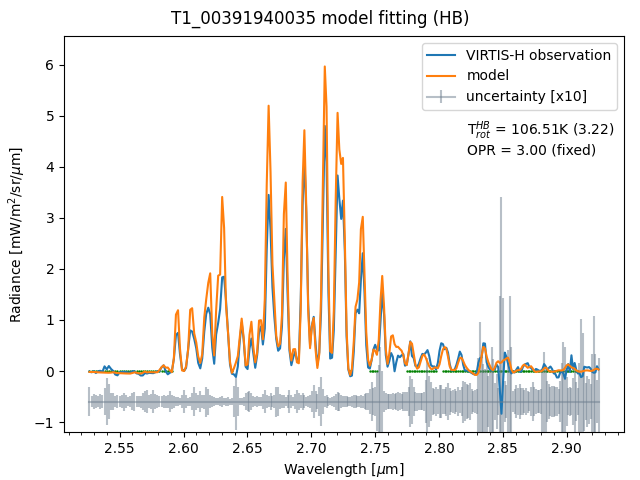}
\end{minipage}
\begin{minipage}{9cm}
\includegraphics[width=9cm]{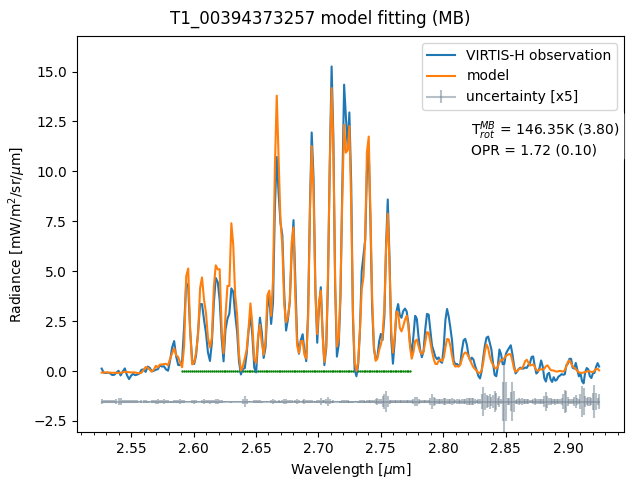}
\end{minipage}\hfill
\begin{minipage}{9cm}
\includegraphics[width=9cm]{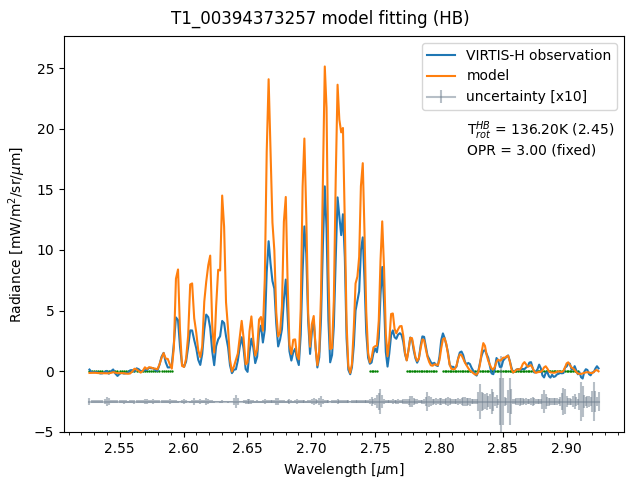}
\end{minipage}
\begin{minipage}{9cm}
\includegraphics[width=9cm]{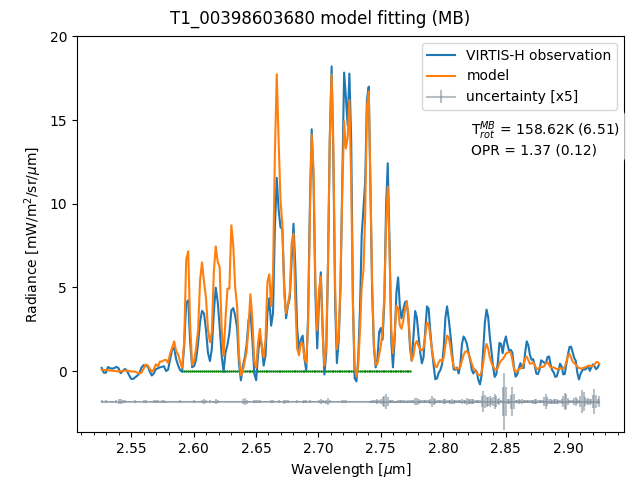}
\end{minipage}\hfill
\begin{minipage}{9cm}
\includegraphics[width=9cm]{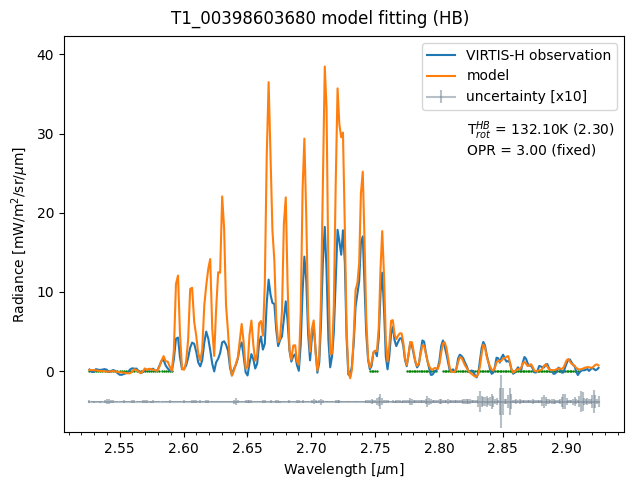}
\end{minipage}
\caption{Spectral fits to the MB (left) and HB (right) spectral region for individual data cubes from dataset S1. See the caption to Figs~\ref{fig:spectrum-MB-H1-H2} and \ref{fig:spectrum-HB-H1-H2}. The observation identification number is given above the plots. Observing and model output parameters are listed in Table ~\ref{tab:S1}.  }
\label{fig:spectrum-MB-HB-S1a}
\end{figure*}

\clearpage

\begin{figure*}[hbt!]
\centering
\begin{minipage}{9cm}
\includegraphics[width=9cm]{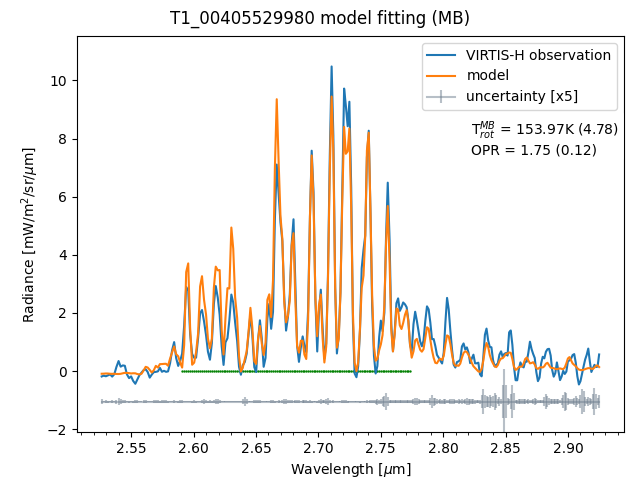}
\end{minipage}\hfill
\begin{minipage}{9cm}
\includegraphics[width=9cm]{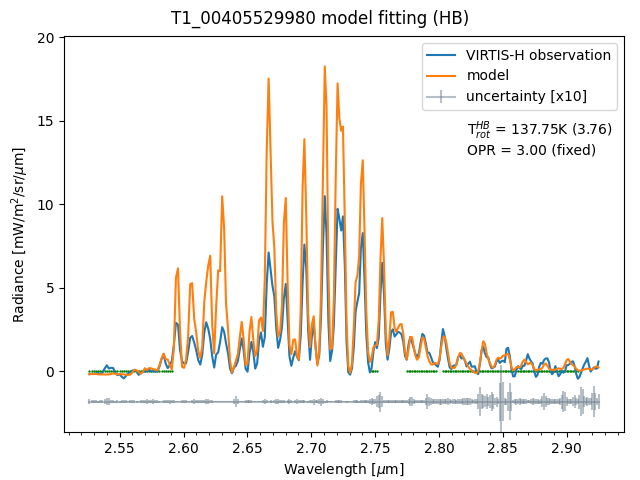}
\end{minipage}
\begin{minipage}{9cm}
\includegraphics[width=9cm]{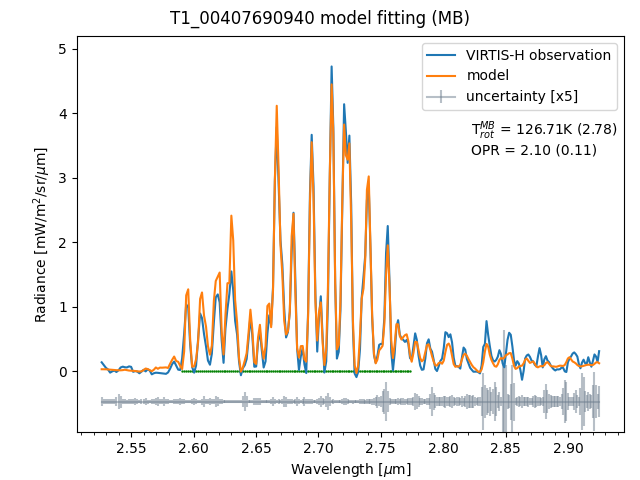}
\end{minipage}\hfill
\begin{minipage}{9cm}
\includegraphics[width=9cm]{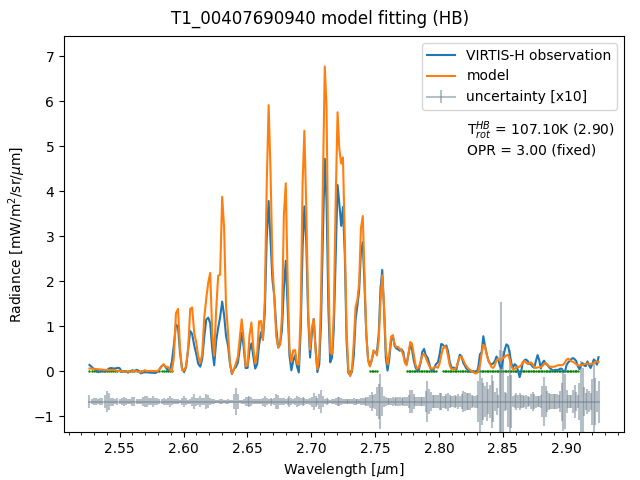}
\end{minipage}
\begin{minipage}{9cm}
\includegraphics[width=9cm]{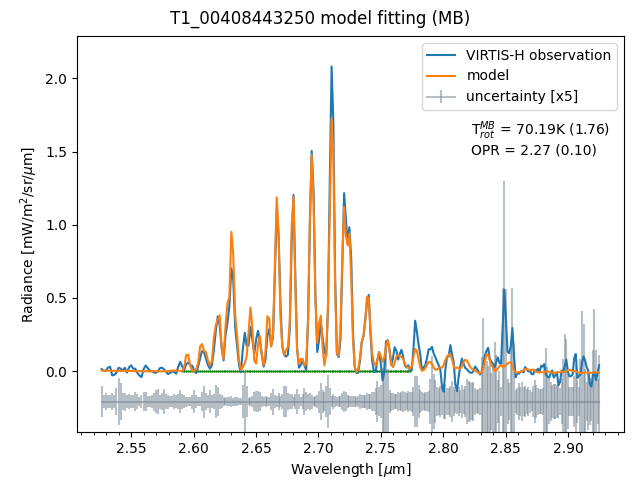}
\end{minipage}\hfill
\begin{minipage}{9cm}
\includegraphics[width=9cm]{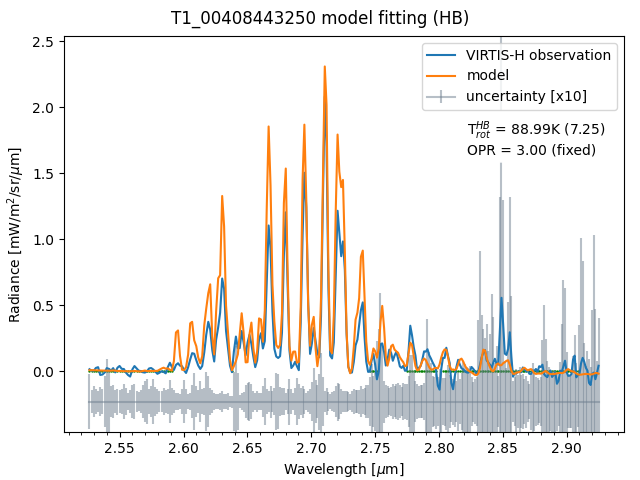}
\end{minipage}
\caption{Spectral fits to the MB (left) and HB (right) spectral regions for individual data cubes from dataset S1. See the captions to Figs~\ref{fig:spectrum-MB-H1-H2} and \ref{fig:spectrum-HB-H1-H2}. The observation identification number (ObsID) is given above the plots. Observing and model output parameters are listed in Table ~\ref{tab:S1}.  }
\label{fig:spectrum-MB-HB-S1b}
\end{figure*}


\section{Results from spectral fits of dataset S1.}
\label{appendix:B}

\onecolumn
\begin{landscape}
\begin{longtable}{lccccccccccc}
\caption{Column densities, rotational temperatures and apparent OPR for individual data cubes (dataset S1).}\label{tab:S1}\\
\hline 
& & & & & &  & & & & &\\
Obs ID & Start time & Int.$^{(a)}$ & $r_{\rm h}$ & $\rho$ $^{(b)}$ & $PA_{\rm LOS}$ $^{(c)}$  & $N_{\rm H_2O}^{\rm MB}$ $^{(d)}$ & $T_{\rm rot}^{\rm MB}$ $^{(d)}$ & $N_{\rm H_2O}^{\rm HB}$ $^{(e)}$ & $T_{\rm rot}^{\rm HB}$ $^{(e)}$ & $f_{\rm atten}$ & OPR$^{(d, f)}$ \\
\noalign{\vskip 2mm}
       & & (h) & (au) & (km) & (deg) & (10$^{20}$ m$^{-2}$) & (K) &  (10$^{20}$ m$^{-2}$) & (K) & & \\
      & & & & & &  & & & & & \\
\hline\noalign{\vskip 2mm} 
\endfirsthead
\hline
& & & & & &  & & & & &\\
Obs ID & Start time & Int.$^{(a)}$ & $r_{\rm h}$ & $\rho$ $^{(b)}$ & $PA_{\rm LOS}$ $^{(c)}$  & $N_{\rm H_2O}^{\rm MB}$ $^{(d)}$ & $T_{\rm rot}^{\rm MB}$ $^{(d)}$ & $N_{\rm H_2O}^{\rm HB}$ $^{(e)}$ & $T_{\rm rot}^{\rm HB}$ $^{(e)}$ & $f_{\rm atten}$ & OPR$^{(d, f)}$ \\
\noalign{\vskip 2mm}
       & & (h) & (au) & (km) & (deg) & (10$^{20}$ m$^{-2}$) & (K) &  (10$^{20}$ m$^{-2}$) & (K) & & \\
              & & & & & &  & & & & &\\
\hline\noalign{\vskip 2mm} 
\endhead
\hline \multicolumn{10}{r}{\textit{Continued on next page}} \\
\endfoot
\hline
\endlastfoot
 T1\_00391940035 & 2015-06-03T08:10:49.4 & 1.27 & 1.506 &  4.18 &  --1.8 &  2.02(0.02) &  98.8(1.8) &  3.33(0.17) & 106.5(3.2) &  0.61(0.03) &  2.16(0.09)\\
 T1\_00391950069 & 2015-06-03T10:58:03.3 & 3.60 & 1.504 &  4.18 &  --2.2 &  2.08(0.03) &  89.1(2.1) &  3.67(0.20) &  93.7(3.9) &  0.57(0.03) &  2.13(0.10)\\
 T1\_00391966054 & 2015-06-03T15:24:28.3 & 4.27 & 1.503 &  4.18 &  --2.6 &  2.01(0.03) &  87.6(1.9) &  3.27(0.15) &  87.1(3.7) &  0.61(0.03) &  2.15(0.10)\\
 T1\_00391986843 & 2015-06-03T21:10:56.7 & 3.60 & 1.502 &  5.18 &  --2.5 &  1.73(0.02) &  78.9(1.5) &  2.83(0.14) &  76.0(4.6) &  0.61(0.03) &  2.11(0.08)\\
 T1\_00392002828 & 2015-06-04T01:37:21.9 & 4.27 & 1.500 &  5.18 &  --2.8 &  1.46(0.02) &  85.6(1.7) &  2.07(0.16) &  88.8(6.1) &  0.71(0.06) &  2.19(0.09)\\
 T1\_00392047400 & 2015-06-04T14:00:13.7 & 2.10 & 1.497 &  3.20 &  --5.9 &  2.45(0.04) & 108.1(2.7) &  4.32(0.31) & 107.3(4.5) &  0.57(0.04) &  2.22(0.13)\\
 T1\_00392477556 & 2015-06-09T13:31:27.3 & 4.21 & 1.466 &  1.92 & --80.9 &  3.27(0.05) & 110.9(2.9) &  5.87(0.30) & 116.8(3.3) &  0.56(0.03) &  1.81(0.10)\\
 T1\_00392813275 & 2015-06-13T10:44:49.8 & 3.60 & 1.443 &  3.00 &  --1.6 &  4.04(0.07) & 116.6(3.2) &  8.40(0.31) & 110.8(2.4) &  0.48(0.02) &  1.91(0.11)\\
 T1\_00393454477 & 2015-06-20T20:51:31.4 & 3.52 & 1.401 &  3.11 &  --1.3 &  4.95(0.08) & 123.7(3.1) &  8.34(0.31) & 120.3(2.3) &  0.59(0.02) &  1.78(0.09)\\
 T1\_00393470172 & 2015-06-21T01:13:06.4 & 4.20 & 1.400 &  3.13 &  --1.1 &  4.45(0.12) & 109.6(4.6) &  8.06(0.62) & 114.3(4.9) &  0.55(0.04) &  1.84(0.16)\\
 T1\_00393490666 & 2015-06-21T06:54:40.3 & 3.52 & 1.399 &  3.15 &  --0.8 &  4.98(0.10) & 116.6(3.4) &  8.56(0.24) & 112.2(1.8) &  0.58(0.02) &  1.82(0.11)\\
 T1\_00393508212 & 2015-06-21T11:47:06.3 & 3.67 & 1.397 &  3.17 &  --0.5 &  4.34(0.07) & 116.3(3.0) &  7.37(0.23) & 113.1(2.0) &  0.59(0.02) &  1.96(0.11)\\
 T1\_00393526856 & 2015-06-21T16:57:50.4 & 3.52 & 1.396 &  3.20 &  --0.2 &  4.74(0.09) & 118.6(3.3) &  8.87(0.26) & 111.3(1.9) &  0.53(0.02) &  1.87(0.11)\\
 T1\_00393874622 & 2015-06-25T17:36:33.2 & 1.95 & 1.375 &  1.65 & --55.4 &  5.32(0.11) & 141.7(4.6) & 10.20(0.54) & 154.4(4.0) &  0.52(0.03) &  1.64(0.12)\\
 T1\_00393911359 & 2015-06-26T03:48:50.1 & 3.63 & 1.373 &  1.78 & --81.1 &  4.33(0.10) & 135.8(5.0) & 10.50(0.76) & 131.2(4.7) &  0.41(0.03) &  1.67(0.13)\\
 T1\_00393927347 & 2015-06-26T08:15:17.7 & 4.31 & 1.372 &  1.93 & --80.8 &  3.73(0.07) & 115.8(3.5) &  7.17(0.41) & 134.5(3.8) &  0.52(0.03) &  1.63(0.10)\\
 T1\_00394131656 & 2015-06-28T16:57:50.7 & 1.95 & 1.361 &  5.28 &  --2.5 &  3.15(0.06) &  93.5(2.5) &  5.53(0.21) &  91.2(2.8) &  0.57(0.02) &  1.78(0.09)\\
 T1\_00394373257 & 2015-07-01T12:00:58.0 & 2.56 & 1.347 &  2.42 &   9.5 &  6.57(0.11) & 146.3(3.8) & 13.40(0.49) & 136.2(2.5) &  0.49(0.02) &  1.72(0.10)\\
 T1\_00394385468 & 2015-07-01T15:24:33.1 & 4.18 & 1.347 &  2.51 &  10.6 &  6.89(0.12) & 144.8(3.9) & 13.10(0.41) & 135.7(2.1) &  0.53(0.02) &  1.68(0.10)\\
 T1\_00394406257 & 2015-07-01T21:10:58.1 & 0.94 & 1.346 &  2.70 &  10.9 &  5.70(0.08) & 134.8(3.0) &  8.77(0.35) & 125.8(2.5) &  0.65(0.03) &  1.84(0.09)\\
 T1\_00394809023 & 2015-07-06T13:03:44.9 & 0.94 & 1.325 &  3.31 &  13.6 &  4.89(0.08) & 121.9(3.2) &  8.51(0.40) & 119.6(3.0) &  0.57(0.03) &  1.76(0.10)\\
 T1\_00395011055 & 2015-07-08T21:11:00.3 & 3.51 & 1.316 &  2.82 & --10.6 &  7.44(0.16) & 131.3(4.5) & 15.00(0.43) & 123.8(1.8) &  0.50(0.02) &  1.61(0.12)\\
 T1\_00395322550 & 2015-07-12T11:46:52.6 & 3.66 & 1.302 &  2.70 &   4.2 &  6.47(0.14) & 140.6(4.8) & 13.30(0.45) & 128.7(2.2) &  0.49(0.02) &  1.53(0.11)\\
 T1\_00395742154 & 2015-07-17T08:15:55.9 & 2.56 & 1.286 &  3.10 &   2.0 &  5.49(0.10) & 123.6(3.6) & 10.10(0.38) & 116.5(2.3) &  0.54(0.02) &  1.75(0.11)\\
 T1\_00396199623 & 2015-07-22T15:23:57.9 & 4.27 & 1.271 &  2.69 &  --3.8 &  7.81(0.17) & 153.0(5.4) & 15.90(0.53) & 138.6(2.3) &  0.49(0.02) &  1.48(0.11)\\
 T1\_00396220410 & 2015-07-22T21:10:24.7 & 3.60 & 1.270 &  2.98 &  --4.1 &  7.96(0.17) & 139.2(4.6) & 16.00(0.42) & 126.3(1.7) &  0.50(0.02) &  1.56(0.11)\\
 T1\_00396659230 & 2015-07-27T23:06:40.5 & 3.54 & 1.259 &  4.01 & --23.4 &  4.80(0.10) & 111.2(3.7) &  8.70(0.27) & 111.5(2.0) &  0.55(0.02) &  1.56(0.10)\\
 T1\_00396711145 & 2015-07-28T13:31:55.4 & 4.22 & 1.258 &  2.01 & --67.5 &  6.64(0.18) & 153.1(6.6) & 17.00(0.90) & 139.3(3.6) &  0.39(0.02) &  1.36(0.12)\\
 T1\_00396731608 & 2015-07-28T19:12:58.9 & 3.54 & 1.257 &  2.12 & --66.7 &  5.39(0.14) & 136.3(5.5) & 12.70(0.56) & 133.8(2.9) &  0.42(0.02) &  1.41(0.12)\\
 T1\_00396826054 & 2015-07-29T21:20:59.8 & 3.24 & 1.255 &  3.56 &  --4.6 &  6.27(0.14) & 127.4(4.5) & 12.40(0.36) & 122.0(1.9) &  0.51(0.02) &  1.60(0.12)\\
 T1\_00396842044 & 2015-07-30T01:47:25.8 & 3.91 & 1.255 &  3.59 &  --4.7 &  6.68(0.15) & 125.0(4.4) & 13.60(0.30) & 111.5(1.4) &  0.49(0.02) &  1.58(0.12)\\
 T1\_00397038715 & 2015-08-01T08:25:16.8 & 0.67 & 1.252 &  3.08 &  --0.2 &  7.10(0.17) & 138.5(5.1) & 13.90(0.46) & 127.7(2.2) &  0.51(0.02) &  1.62(0.13)\\
 T1\_00397849779 & 2015-08-10T17:35:55.8 & 3.82 & 1.244 &  6.10 &  15.4 &  4.60(0.10) & 110.2(3.8) & 10.20(0.31) &  96.1(2.1) &  0.45(0.02) &  1.49(0.11)\\
 T1\_00397871165 & 2015-08-10T23:32:29.0 & 2.80 & 1.244 &  6.23 &  16.3 &  4.80(0.10) & 105.5(3.5) &  9.43(0.30) &  93.1(2.3) &  0.51(0.02) &  1.66(0.11)\\
 T1\_00397885968 & 2015-08-11T03:39:04.9 & 3.82 & 1.243 &  6.35 &  17.0 &  4.56(0.11) & 105.3(3.9) &  9.38(0.37) &  94.5(2.8) &  0.49(0.02) &  1.49(0.11)\\
 T1\_00398347798 & 2015-08-16T11:56:25.1 & 3.31 & 1.244 &  3.58 &  --3.4 &  7.81(0.19) & 145.8(5.7) & 18.30(0.49) & 112.1(1.7) &  0.43(0.02) &  1.39(0.12)\\
 T1\_00398603680 & 2015-08-19T11:08:06.5 & 3.24 & 1.246 &  2.96 &  --7.1 &  7.93(0.20) & 158.6(6.5) & 17.40(0.61) & 132.1(2.3) &  0.46(0.02) &  1.37(0.12)\\
 T1\_00398619671 & 2015-08-19T15:34:33.4 & 3.91 & 1.246 &  2.97 &  --8.4 &  8.25(0.20) & 147.5(5.7) & 17.20(0.50) & 127.3(1.9) &  0.48(0.02) &  1.47(0.12)\\
 T1\_00398640452 & 2015-08-19T21:20:58.5 & 3.24 & 1.246 &  3.47 &  --8.5 &  7.07(0.18) & 145.2(6.0) & 16.40(0.61) & 126.5(2.4) &  0.43(0.02) &  1.44(0.12)\\
 T1\_00398729980 & 2015-08-20T22:13:06.6 & 3.91 & 1.247 &  4.58 & --12.5 &  6.60(0.16) & 128.8(4.7) & 14.80(0.45) & 106.0(2.0) &  0.45(0.02) &  1.41(0.11)\\
 T1\_00398970868 & 2015-08-23T17:07:54.5 & 3.10 & 1.250 &  3.78 & --13.9 &  7.41(0.18) & 129.4(4.9) & 15.80(0.47) & 115.3(1.9) &  0.47(0.02) &  1.55(0.12)\\
 T1\_00398986563 & 2015-08-23T21:29:29.5 & 3.78 & 1.250 &  3.82 & --14.9 &  7.52(0.19) & 136.2(5.4) & 17.40(0.43) & 114.4(1.5) &  0.43(0.02) &  1.46(0.12)\\
 T1\_00399198695 & 2015-08-26T08:17:52.6 & 1.02 & 1.254 &  4.25 &  --6.2 &  6.13(0.16) & 125.0(4.9) & 13.20(0.45) & 110.4(2.2) &  0.46(0.02) &  1.53(0.12)\\
 T1\_00399208316 & 2015-08-26T10:58:13.4 & 3.56 & 1.254 &  3.95 &  --7.2 &  7.21(0.18) & 128.6(4.9) & 14.80(0.48) & 112.0(2.1) &  0.49(0.02) &  1.45(0.11)\\
 T1\_00399684040 & 2015-08-31T23:06:58.1 & 3.31 & 1.265 &  5.64 & --34.7 &  5.13(0.12) & 111.2(4.0) & 11.20(0.24) &  96.6(1.5) &  0.46(0.01) &  1.66(0.12)\\
 T1\_00399699731 & 2015-09-01T03:28:32.2 & 4.07 & 1.265 &  5.74 & --35.8 &  4.08(0.08) & 111.2(3.5) &  8.26(0.27) &  91.0(2.4) &  0.49(0.02) &  1.58(0.10)\\
 T1\_00399720221 & 2015-09-01T09:10:08.7 & 3.56 & 1.266 &  5.87 & --37.0 &  4.44(0.10) & 109.2(3.7) &  9.46(0.30) &  90.6(2.4) &  0.47(0.02) &  1.70(0.12)\\
 T1\_00400407713 & 2015-09-09T08:15:16.0 & 1.08 & 1.288 &  3.14 &  --6.6 &  6.15(0.18) & 165.3(7.6) & 15.60(0.96) & 121.9(4.0) &  0.39(0.03) &  1.32(0.13)\\
 T1\_00400433767 & 2015-09-09T15:29:34.0 & 4.05 & 1.289 &  3.72 &  --7.7 &  6.37(0.15) & 148.2(5.6) & 14.30(0.40) & 111.8(1.8) &  0.45(0.02) &  1.42(0.12)\\
 T1\_00400748481 & 2015-09-13T06:54:44.2 & 3.51 & 1.301 &  3.58 &  --4.8 &  6.20(0.14) & 140.4(5.0) & 13.70(0.33) & 114.4(1.5) &  0.45(0.02) &  1.51(0.12)\\
 T1\_00400765976 & 2015-09-13T11:46:19.3 & 3.64 & 1.302 &  3.61 &  --5.5 &  6.83(0.18) & 145.0(6.0) & 16.20(0.49) & 117.0(1.9) &  0.42(0.02) &  1.38(0.12)\\
 T1\_00401664583 & 2015-09-23T21:16:05.4 & 3.31 & 1.344 &  4.65 &   4.0 &  5.32(0.11) & 115.0(3.7) & 11.00(0.36) &  98.7(2.3) &  0.48(0.02) &  1.61(0.11)\\
 T1\_00401701357 & 2015-09-24T07:28:56.0 & 3.31 & 1.346 &  5.05 &   4.9 &  4.46(0.09) & 105.3(3.5) &  9.21(0.29) &  95.8(2.3) &  0.48(0.02) &  1.66(0.11)\\
 T1\_00401717336 & 2015-09-24T11:55:21.3 & 4.07 & 1.347 &  5.02 &   5.5 &  4.62(0.09) & 108.7(3.4) &  9.66(0.32) &  99.1(2.3) &  0.48(0.02) &  1.59(0.10)\\
 T1\_00401738122 & 2015-09-24T17:41:48.0 & 3.31 & 1.348 &  5.48 &   5.6 &  3.57(0.07) & 106.5(3.1) &  6.83(0.28) &  91.6(3.1) &  0.52(0.02) &  1.63(0.09)\\
 T1\_00401975571 & 2015-09-27T11:46:18.5 & 3.64 & 1.361 &  6.79 &  10.9 &  2.98(0.05) &  96.9(2.7) &  5.92(0.19) &  89.1(2.4) &  0.50(0.02) &  1.66(0.09)\\
 T1\_00401994267 & 2015-09-27T16:57:54.5 & 3.51 & 1.362 &  6.88 &  11.5 &  2.56(0.05) &  92.2(2.7) &  4.73(0.21) &  78.3(3.8) &  0.54(0.03) &  1.65(0.09)\\
 T1\_00402009960 & 2015-09-27T21:19:27.5 & 4.18 & 1.363 &  7.13 &  11.8 &  2.56(0.05) &  90.6(2.6) &  4.63(0.21) &  80.2(3.8) &  0.55(0.03) &  1.74(0.09)\\
 T1\_00402963204 & 2015-10-08T21:59:43.4 & 4.33 & 1.422 &  2.66 & --73.8 &  5.41(0.12) & 119.8(4.0) & 10.50(0.39) & 118.5(2.3) &  0.52(0.02) &  1.64(0.12)\\
 T1\_00402983991 & 2015-10-09T03:46:10.0 & 3.56 & 1.423 &  2.89 & --73.3 &  5.62(0.14) & 120.4(4.5) & 10.90(0.41) & 109.7(2.4) &  0.52(0.02) &  1.64(0.13)\\
 T1\_00403020756 & 2015-10-09T13:59:01.5 & 3.56 & 1.425 &  3.34 & --72.5 &  5.50(0.12) & 110.4(3.6) & 11.40(0.41) & 102.0(2.5) &  0.48(0.02) &  1.54(0.10)\\
 T1\_00403431934 & 2015-10-14T08:11:53.0 & 1.02 & 1.453 &  2.93 &   4.1 &  5.99(0.11) & 134.6(3.8) & 11.70(0.49) & 122.7(2.7) &  0.51(0.02) &  1.71(0.10)\\
 T1\_00403604772 & 2015-10-16T08:12:34.7 & 4.33 & 1.465 &  4.99 &   7.6 &  3.43(0.06) &  94.1(2.5) &  5.16(0.33) &  77.0(5.5) &  0.66(0.04) &  1.67(0.09)\\
 T1\_00403625558 & 2015-10-16T13:59:00.8 & 3.56 & 1.467 &  5.00 &   8.5 &  3.83(0.07) &  98.7(2.8) &  5.85(0.43) &  75.6(6.5) &  0.65(0.05) &  1.66(0.09)\\
 T1\_00403641546 & 2015-10-16T18:25:25.9 & 4.33 & 1.468 &  5.02 &   9.4 &  3.81(0.07) &  98.5(2.7) &  6.11(0.33) &  72.4(5.0) &  0.62(0.04) &  1.70(0.09)\\
 T1\_00404036434 & 2015-10-21T08:06:53.1 & 1.27 & 1.496 &  3.34 &   2.5 &  5.74(0.11) & 118.4(3.5) & 10.60(0.55) & 118.3(3.3) &  0.54(0.03) &  1.68(0.10)\\
 T1\_00404046803 & 2015-10-21T10:59:42.6 & 3.56 & 1.497 &  3.46 &   3.0 &  4.44(0.10) & 126.4(4.5) &  9.01(0.45) & 105.5(3.2) &  0.49(0.03) &  1.66(0.12)\\
 T1\_00404062791 & 2015-10-21T15:26:10.2 & 4.07 & 1.499 &  3.48 &   3.8 &  4.90(0.11) & 114.7(3.8) &  9.71(0.39) & 100.5(2.7) &  0.50(0.02) &  1.62(0.11)\\
 T1\_00404172794 & 2015-10-22T21:59:43.5 & 4.33 & 1.507 &  4.31 &   9.0 &  3.85(0.07) & 100.9(3.0) &  6.98(0.38) &  93.3(3.9) &  0.55(0.03) &  1.70(0.10)\\
 T1\_00404193581 & 2015-10-23T03:46:10.2 & 3.56 & 1.508 &  4.33 &  10.2 &  3.55(0.07) &  98.1(2.9) &  5.95(0.50) &  96.2(5.8) &  0.60(0.05) &  1.79(0.11)\\
 T1\_00404209572 & 2015-10-23T08:12:34.7 & 4.33 & 1.509 &  4.35 &  11.4 &  4.14(0.08) & 101.0(3.1) &  6.91(0.36) & 103.2(3.4) &  0.60(0.03) &  1.73(0.10)\\
 T1\_00404486062 & 2015-10-26T13:00:47.9 & 1.78 & 1.530 &  1.99 &  27.2 &  6.66(0.12) & 153.2(4.3) & 10.00(0.70) & 149.2(5.1) &  0.67(0.05) &  1.55(0.09)\\
 T1\_00404900299 & 2015-10-31T08:11:44.1 & 1.08 & 1.562 &  3.14 &  --0.8 &  3.66(0.08) & 110.1(3.6) &  6.10(0.32) & 121.1(3.3) &  0.60(0.03) &  1.61(0.11)\\
 T1\_00405245595 & 2015-11-04T08:10:12.3 & 1.20 & 1.589 &  3.71 &   3.6 &  2.81(0.06) &  98.4(3.2) &  4.79(0.27) &  93.5(4.1) &  0.59(0.04) &  1.83(0.12)\\
 T1\_00405382267 & 2015-11-05T22:08:04.3 & 4.12 & 1.601 &  4.86 &   7.6 &  2.44(0.04) &  88.5(2.2) &  4.12(0.16) &  90.8(2.9) &  0.59(0.02) &  1.76(0.08)\\
 T1\_00405403052 & 2015-11-06T03:54:29.3 & 3.45 & 1.602 &  4.84 &   8.6 &  2.12(0.04) &  80.4(2.1) &  3.69(0.15) &  83.2(3.4) &  0.57(0.03) &  1.83(0.09)\\
 T1\_00405419039 & 2015-11-06T08:20:56.2 & 4.12 & 1.604 &  5.22 &   8.9 &  2.04(0.04) &  80.5(2.3) &  3.49(0.12) &  86.5(2.8) &  0.58(0.02) &  1.84(0.09)\\
 T1\_00405505096 & 2015-11-07T08:15:13.2 & 1.05 & 1.610 &  1.83 & --32.5 &  6.06(0.11) & 168.8(4.8) & 13.40(0.88) & 152.9(5.0) &  0.45(0.03) &  1.80(0.11)\\
 T1\_00405529980 & 2015-11-07T15:09:57.3 & 4.20 & 1.613 &  1.88 & --34.5 &  6.58(0.13) & 154.0(4.8) & 14.00(0.79) & 137.8(3.8) &  0.47(0.03) &  1.75(0.12)\\
 T1\_00405604159 & 2015-11-08T11:46:16.1 & 3.67 & 1.619 &  2.44 & --10.9 &  5.38(0.09) & 129.5(3.2) & 10.30(0.35) & 118.0(2.2) &  0.52(0.02) &  1.69(0.09)\\
 T1\_00405622853 & 2015-11-08T16:57:50.4 & 3.52 & 1.620 &  2.66 & --10.4 &  5.01(0.09) & 128.9(3.8) &  9.15(0.46) & 119.7(3.2) &  0.55(0.03) &  1.65(0.10)\\
 T1\_00405638548 & 2015-11-08T21:19:25.4 & 4.20 & 1.622 &  2.86 &  --9.7 &  4.77(0.09) & 122.7(3.4) &  8.37(0.34) & 113.1(2.6) &  0.57(0.03) &  1.78(0.10)\\
 T1\_00405747416 & 2015-11-10T03:33:53.4 & 4.05 & 1.631 &  6.16 &  --4.1 &  1.91(0.04) &  76.9(2.2) &  2.90(0.21) &  60.3(7.3) &  0.66(0.05) &  1.87(0.10)\\
 T1\_00405767912 & 2015-11-10T09:15:29.3 & 3.37 & 1.632 &  6.30 &  --2.9 &  2.06(0.04) &  75.2(2.4) &  3.55(0.24) &  63.0(6.8) &  0.58(0.04) &  1.78(0.10)\\
 T1\_00406109599 & 2015-11-14T08:06:44.5 & 1.08 & 1.660 &  3.58 &   0.1 &  2.71(0.05) & 100.0(2.7) &  4.83(0.27) &  82.6(4.6) &  0.56(0.03) &  1.93(0.10)\\
 T1\_00406119388 & 2015-11-14T10:49:53.4 & 3.37 & 1.661 &  4.08 &   0.3 &  2.63(0.05) &  91.4(2.6) &  4.64(0.25) &  97.4(3.9) &  0.57(0.03) &  1.82(0.10)\\
 T1\_00407060001 & 2015-11-25T08:06:46.4 & 1.35 & 1.741 &  2.93 &  --2.8 &  3.88(0.07) & 115.6(3.0) &  6.54(0.31) & 110.6(3.1) &  0.59(0.03) &  1.94(0.11)\\
 T1\_00407233296 & 2015-11-27T08:12:27.6 & 4.30 & 1.756 &  3.01 & --13.0 &  3.33(0.06) & 110.6(2.9) &  4.59(0.28) &  90.9(4.6) &  0.73(0.05) &  1.75(0.09)\\
 T1\_00407254083 & 2015-11-27T13:58:55.2 & 3.40 & 1.758 &  3.10 & --13.7 &  3.27(0.06) & 103.8(2.8) &  4.30(0.39) &  98.8(6.2) &  0.76(0.07) &  1.74(0.09)\\
 T1\_00407270065 & 2015-11-27T18:25:19.5 & 4.30 & 1.759 &  3.19 & --14.3 &  2.93(0.05) &  99.2(2.7) &  3.77(0.33) &  64.8(8.9) &  0.78(0.07) &  1.81(0.10)\\
 T1\_00407664582 & 2015-12-02T08:06:40.7 & 1.35 & 1.793 &  2.48 &  --5.0 &  2.95(0.04) & 123.0(2.8) &  4.64(0.27) & 117.3(3.6) &  0.64(0.04) &  2.03(0.10)\\
 T1\_00407674953 & 2015-12-02T10:59:31.7 & 3.52 & 1.794 &  2.48 &  --5.8 &  3.15(0.05) & 121.4(2.8) &  4.53(0.24) & 126.0(3.4) &  0.70(0.04) &  2.01(0.10)\\
 T1\_00407690940 & 2015-12-02T15:25:58.7 & 4.20 & 1.796 &  2.49 &  --6.8 &  3.23(0.05) & 126.7(2.8) &  5.43(0.24) & 107.1(2.9) &  0.59(0.03) &  2.10(0.11)\\
 T1\_00407711726 & 2015-12-02T21:12:24.5 & 3.52 & 1.798 &  2.50 &  --8.3 &  3.20(0.05) & 133.1(3.3) &  5.32(0.22) & 119.9(2.6) &  0.60(0.03) &  2.01(0.12)\\
 T1\_00408406055 & 2015-12-10T22:04:33.4 & 4.20 & 1.859 &  2.84 &  --2.5 &  2.48(0.04) & 104.3(2.4) &  3.90(0.21) & 109.5(3.4) &  0.64(0.04) &  2.09(0.11)\\
 T1\_00408443250 & 2015-12-11T08:20:59.6 & 4.05 & 1.862 &  5.67 & --46.1 &  0.92(0.01) &  70.2(1.8) &  1.73(0.17) &  89.0(7.3) &  0.53(0.05) &  2.27(0.10)\\
 T1\_00409143171 & 2015-12-19T10:49:50.1 & 3.45 & 1.924 &  2.80 &  --0.5 &  2.12(0.03) & 110.1(2.4) &  3.48(0.27) & 107.4(5.1) &  0.61(0.05) &  2.34(0.12)\\
 T1\_00409158866 & 2015-12-19T15:11:25.2 & 4.12 & 1.925 &  2.90 &  --0.4 &  2.18(0.03) & 104.3(2.3) &  3.39(0.22) &  91.3(5.0) &  0.64(0.04) &  2.29(0.12)\\
 T1\_00410094153 & 2015-12-30T10:59:32.4 & 3.52 & 2.009 &  2.75 &   1.7 &  1.91(0.03) & 108.9(2.3) &  3.53(0.25) & 100.1(4.7) &  0.54(0.04) &  2.38(0.12)\\
 T1\_00410110139 & 2015-12-30T15:25:58.5 & 4.20 & 2.010 &  2.80 &   2.0 &  1.77(0.02) & 106.4(2.2) &  2.63(0.24) & 108.5(5.9) &  0.67(0.06) &  2.32(0.11)\\
 T1\_00410130927 & 2015-12-30T21:12:26.0 & 3.52 & 2.012 &  2.89 &   2.5 &  2.00(0.03) & 109.6(2.3) &  3.33(0.24) &  96.4(4.9) &  0.60(0.04) &  2.28(0.11)\\
 T1\_00411340224 & 2016-01-13T21:07:23.9 & 3.67 & 2.120 &  2.67 &   0.3 &  1.33(0.02) & 109.6(2.4) &  2.17(0.21) &  94.5(6.9) &  0.61(0.06) &  2.72(0.15)\\     

\hline
\end{longtable}
\begin{minipage}{0.93\linewidth}
$^{(a)}$ Total duration of cube acquisition.

$^{(b)}$ Azimuth position angle of LOS with respect to the comet-Sun line (anti-clockwise).

$^{(c)}$ Mean distance of FOV to comet centre.  

$^{(d)}$ From spectral fitting of the MB spectral region (affected by optical depth effects); uncertainties in parentheses.

$^{(e)}$ From spectral fitting of the HB spectral region (representative of the coma); uncertainties in parentheses.

$^{(f)}$ Apparent OPR.

This table is available in ASCII format: \url{https://zenodo.org/record/6303110#.YhyljCbjI34}
\end{minipage}
\end{landscape}
\twocolumn

\end{appendix}

\end{document}